%% file: SPL-v6.tex
\documentclass[12pt, draftclsnofoot, onecolumn]{IEEEtran}

\usepackage{bbm}
\usepackage{bbold}
\usepackage{amsmath}
\usepackage{amssymb}
\usepackage[dvips]{graphicx}
\usepackage{graphicx}
\usepackage{cite}
\usepackage{epsfig}
\usepackage{bigints}
\usepackage{stfloats}
\usepackage{amsthm}
\usepackage{float}
\usepackage{relsize}
\usepackage{bbold}
\usepackage{bbm, dsfont}
\usepackage{xcolor}
\usepackage{mathrsfs}
\usepackage{booktabs}
\usepackage{physics}
\usepackage{makecell}
\usepackage{caption}
\usepackage[linesnumbered,ruled,vlined]{algorithm2e}

\newtheorem{lemma}{Lemma}

\newtheorem{theorem}{Theorem}

\newcommand{\chosenAction}{a}
\newcommand{\packetArrRate}{A}
\newcommand{\actionSet}{\mathcal{A}}

\newcommand{\matchPolicy}{b}
\newcommand{\contDelay}{c}

\newcommand{\sbsClusterSet}{\mathcal{C}}
\newcommand{\contDemand}{D}
\newcommand{\cacheSize}{d}
\newcommand{\rbRepresentation}{e}

\newcommand{\contCatalog}{F}
\newcommand{\contCatalogSet}{\mathcal{F}}
\newcommand{\gibbDist}{G}
\newcommand{\rbsArrRate}{h}
\newcommand{\rbQueueEvo}{H}
\newcommand{\propCachePolicy}{I}
\newcommand{\cost}{J}
\newcommand{\noOfContentClasses}{K}
\newcommand{\contentClassesSet}{\mathcal{K}}
\newcommand{\numBitsPerPacket}{l_p}
\newcommand{\lyapunov}{L}

\newcommand{\rbsSize}{M}
\newcommand{\rbsSet}{\mathcal{M}}
\newcommand{\numCacheUp}{N}
\newcommand{\ueCoverage}{\mathcal{N}}
\newcommand{\txPower}{p}
\newcommand{\popularity}{P}
\newcommand{\ueRequests}{q}
\newcommand{\ueReqDist}{\mathbbm{q}}

\newcommand{\dataQueue}{Q}
\newcommand{\rate}{R}
\newcommand{\regret}{r}
\newcommand{\sbsSize}{S}
\newcommand{\sbsSet}{\mathcal{S}}
\newcommand{\qosReq}{T}
\newcommand{\sbsClustTime}{T_1}
\newcommand{\cacheUpdateTime}{T_2}
\newcommand{\uePrefTime}{T_{\mathrm{FIX}}}
\newcommand{\utility}{u}
\newcommand{\ueSize}{U}
\newcommand{\ueSet}{\mathcal{U}}
\newcommand{\ueRbsDefVec}{v}
\newcommand{\utilDelayTradeoff}{V}
\newcommand{\transBits}{W}
\newcommand{\frthaulCap}{X}
\newcommand{\schedVar}{Y}
\newcommand{\sysStateSmall}{z}
\newcommand{\sysStateBig}{Z}
\newcommand{\sysStateSet}{\mathcal{Z}}

\newcommand{\sbsDensity}{\lambda_{\mathrm{SBS}}}
\newcommand{\ueDensity}{\lambda_{\mathrm{UE}}}
\newcommand{\packetSize}{\mu}
\newcommand{\insCache}{\Xi}
\newcommand{\channel}{\Psi}
\newcommand{\bwPerRB}{\omega}
\newcommand{\sbsTransDur}{\alpha}
\newcommand{\locGlobTradeof}{\beta}
\newcommand{\frthaulPerUe}{\nu}
\newcommand{\pwrLawExp}{\delta}
\newcommand{\aggCost}{\Upsilon}
\newcommand{\noise}{\sigma^2}
\newcommand{\ImpPopSim}{\sigma_c^2}
\newcommand{\learningPop}{\pi}
\newcommand{\cacheUpTime}{\tau}
\newcommand{\CacheTempCoeff}{\xi}
\newcommand{\learParam}{\gamma}
\newcommand{\virtQueue}{\Gamma}
\newcommand{\realMatching}{\delta}
\newcommand{\realScheduling}{\psi}
\newcommand{\PopPowerLawExp}{\varphi}

\theoremstyle{definition}
\newtheorem{definition}{Definition}

\IEEEoverridecommandlockouts

\begin{document}

\title{Latency-Aware Radio Resource Optimization in Learning-based Cloud-Aided Small Cell Wireless Networks} %
\author{	
	\IEEEauthorblockN{Syed Tamoor-ul-Hassan, Sumudu Samarakoon, Mehdi Bennis and Matti Latva-aho} \\
	\IEEEauthorblockA{ Center for Wireless Communications, University of Oulu, Finland} \\
	Email: \{tamoor-ul-hassan.syed, sumudu.samarakoon, mehdi.bennis, matti.latva-aho\}@oulu.fi
\vspace{-5ex} }
\providecommand{\keywords}[1]{\textbf{\textit{Index terms---}} #1}
\vspace{-4ex}
\maketitle

\begin{abstract}
Low latency communication is one of the fundamental requirements for 5G wireless networks and beyond. In this paper, a novel approach for joint caching, user scheduling and resource allocation is proposed for minimizing the queuing latency in serving users' requests in cloud-aided wireless networks. Due to the slow temporal variations in user requests, a time-scale separation technique is used to decouple the joint caching problem from user scheduling and radio resource allocation problems. To serve the spatio-temporal user requests under storage limitations, a Reinforcement Learning (RL) approach is used to optimize the caching strategy at the small cell base stations by minimizing the content fetching cost. A spectral clustering algorithm is proposed to speed-up the convergence of the RL algorithm for a large content caching problem by clustering contents based on user requests. Meanwhile, a dynamic mechanism is proposed to locally group coupled base stations based on user requests to collaboratively optimize the caching strategies. To further improve the latency in fetching and serving user requests, a dynamic matching algorithm is proposed to schedule users and to allocate users to radio resources based on user requests and queue lengths under probabilistic latency constraints. Simulation results show the proposed approach significantly reduces the average delay from 21\% to 90\% compared to random caching strategy, random resource allocation and random scheduling baselines.
\end{abstract}
\section{Introduction}
\label{sec:Intro}
The proliferation of mobile data traffic and smart wireless devices has significantly reshaped the landscape of wireless content networks \cite{Ref1_Intro}. Taming the data tsunami requires rethinking the current network architecture to ensure efficient bandwidth usage. At the same time, efficient processing supporting low latency transmission of tactile multimedia such as real time interactive hologram services, immersive media and high definition multimedia streaming services are key cornerstone for 5G networks and beyond. Edge caching has been envisioned as one of the key enabling technologies to overcome the challenges associated with next generation networks among others \cite{Ref4_Intro}\cite{Ref10_Intro}. Apart from improving users' Quality-of-Experience (QoE), edge caching significantly offloads different network entities including radio access network, core network and backhaul, easing network congestion and improving latency and energy efficiency \cite{Ref11_Intro, Ref13_Intro, Ref17_Intro}.  \\
\indent Motivated by the inherent user behavior, \textit{Machine Learning} (ML) driven caching recently received significant attention in the context of cellular networks \cite{Ref18_CacheAI,Ref18_CacheAI1,Ref18_CacheAI2}. ML allows to learn users' statistical behavior and subsequently optimize caching operation \cite{Ref18_AI} \cite{Ref18_AI2}. 
\cite{Ref18_CacheAI, Ref18_CacheAI1, Ref18_CacheAI2} ignore the impact of fronthaul capacity while relying on centralized cloud server to optimize the caching policy. Worth noting is that centralized ML algorithms require sufficient training data samples thereby increasing the delay which undermines latency-sensitive applications. 
While \cite{Ref18_CacheAI4,Ref18_CacheAI5,Ref18_CacheAI6} incorporate the fronthaul capacity requirements, these works consider fixed content popularity, and ignore the dynamic arrival of user requests and neglect latency constraints. \\ 
\indent Due to the stringent latency requirements and spatio-temporal user behavior, \textit{edge ML} has been proposed to address the shortcomings of cloud-based ML algorithms \cite{Ref19_AIEdge1}\cite{Ref19_AIEdge2}. Instead of training at a centralized cloud server, \textit{edge ML} exploits learning over distributed nodes to learn the global behavior and  provide services at the edge \cite{Ref18_CacheEdge, Ref18_CacheEdge1}. 
As shown in prior works \cite{Ref18_CacheEdge, Ref18_CacheEdge1, Ref18_CollaborativeAI2, Ref18_CacheEdge3}, big-data-enabled ML algorithms require high computation resources that may not be always available at the edge devices. Rather than relying on a huge data set, edge devices may consider a subset of the big data to train the ML algorithms \cite{Ref18_CacheEdge4}. However, data sparsity and limited computation resources may affect the performance of decentralized ML algorithms \cite{Ref18_CacheEdge4}. Therefore, a hybrid architecture integrating both the benefits of centralized and distributed learning has been proposed to enable ML in a collaborative fashion \cite{Ref18_CollaborativeAI1, Ref18_CollaborativeAI3}. However, these works optimize throughput and latency while ignoring the content caching problem.  \\
\indent Within the context of caching, \cite{Ref18_CollaborativeCacheEdge1, Ref18_CollaborativeCacheEdge3, Ref18_CollaborativeCacheEdge4} recently showed the benefits of collaborative cloud and edge caching. Although \cite{Ref18_CollaborativeCacheEdge1} optimize the caching policy in a cloud-based wireless network without employing ML, the work does not explore the tradeoff between cloud/edge based caching policy and their inherent local and global content popularity dynamics. Further, the impact of fronthaul has not been considered. The tradeoff between cloud/edge based caching policy using RL has been considered in \cite{Ref18_CollaborativeCacheEdge3} \cite{Ref18_CollaborativeCacheEdge4}. However the framework restricts the number of cache updates, does not assume collaboration between small cell base stations (SBSs) and ignores the impact of radio resource management and user scheduling. Besides these shortcomings, the work also ignores latency and reliability constraints. In terms of radio resource management and user scheduling, the works in \cite{Ref23_Intro, Ref24_Intro, Ref1:Lyapunov_DPP} consider the dynamic queue state information (QSI) and channel state information (CSI) where the base station schedules a user based on a fixed caching policy. Furthermore, these studies do not consider adaptive cache policies and dynamics of request distribution. In practice, due to the dynamic CSI, QSI and requests arrival, there is a need to jointly address caching and user scheduling problems. In \cite{Ref25_Intro}, the authors propose a power control algorithm and cache control algorithm to satisfy users' quality-of-service (QoS) requirements. However, the proposed algorithms rely on a centralized controller which is inapplicable for ultra dense network where coordination imposes a huge network overhead. Meanwhile, \cite{Ref26_Intro, Ref27_Intro, Ref28_Intro_2} proposes RL-based caching strategies based on static channel conditions and ignoring stringent QoS requirements. \\
\indent The main contribution of this paper is to propose a novel joint adaptive caching policy and user scheduling mechanism for wireless small cell networks under limited cache constraints. In the studied model, the objective of each SBS is to minimize the latency of serving users' request while satisfying their QoE. The problem is addressed leveraging two separate timescales where the cache policy is optimized in a long timescale, while user scheduling and resource allocation are carried out in a short time scale. In a long time scale, each SBS learns the content popularity profiles with the aid of regret learning \cite{Ref1:Reinforcement_Learning} and uses it to optimize the cache policy in a decentralized manner. For faster convergence, regret learning is applied by classifying contents into multiple classes such that contents of a given class have similar demands. Based on the popularity profile, each SBS updates its cache. For user scheduling, each SBS dynamically allocates resources to a given task based on the resource-task matching queuing approach \cite{Ref29_Intro}. Accounting the random request arrivals and dynamic channel conditions, resource management and user scheduling are cast as a stochastic network optimization problem \cite{Ref26_Intro_Stochastic}. 
In order to maximize network utility while ensuring queue stability, user scheduling via matching is solved using the \textit{Drift-Plus-Penalty} (DPP)-based lyapunov optimization framework \cite{Ref30_Intro}. Due to the coupling of radio resources among SBSs, the problem becomes a difference of convex programming problem and it is solved using a local heuristic \cite{Ref1:CCP}. The convergence and the complexity of the proposed algorithm are analyzed and validated by simulations. Simulation results shows upto 90\% reduction in latency of the proposed caching and dynamic matching algorithms compared to fixed caching policy and random matching algorithm. \\
\indent The rest of the paper is organized as follows. In Section II, the system model is described including the user request model and the queueing model. The problem formulation are presented in Section III. In Section IV, clustering method for both SBS and contents is presented. Section V describes the reinforcement learning method to track content popularity, while Section VI presents the lyapunov optimization framework for user scheduling and resource allocation. Finally, Section VII describes the performance of the proposed scheme with baseline methods followed by conclusions in Section VIII.
\begin{table} 
\caption{List of Symbols (subscript $i$ and $j$ represents SBSs and users respectively)}
\vspace{-0.2cm}
\begin{minipage}{0.5\textwidth}
\begin{tabular}{|p{1.0cm} p{6.40cm}|}
\hline
\textbf{Symbols} & \textbf{Description} \\
\hline
$\sbsSize$,$\ \sbsSet$ & Size, set of SBSs \\
$\ueSet$,$\ \ueSize$ & Size, set of users \\
$\sbsDensity$,$\ \ueDensity$ & Density of SBSs, UEs \\
$\packetArrRate_{ji}^f$ & Number of packets of file $f$ at $j$ destined for $i$ \\
$\actionSet_i$ & Action space of $i$ \\ 
$\chosenAction_s(t)$ & Action of SBS $s$ at time $t$ \\
$\boldsymbol{\insCache}_s(t)$ & Cache of SBS $s$ at time $t$ \\
$\matchPolicy_{su}^m(t)$ & Allocation of RB $m$ to user $u$ by SBS $s$ at time $t$ \\
$\sbsClusterSet$ & SBSs cluster set \\
$\cacheSize$ & Cache size per SBS \\
$\boldsymbol{\contDemand}_s$ & Content demand vector at SBS $s$ \\
$\contCatalog$, $\contCatalogSet$ & Size, set of content catalog \\
$\boldsymbol{\gibbDist}_i$ & Boltzman Gibbs Distribution vector at $i$ \\
$\rbsArrRate_s(t)$ & Arrival rate of RBs for SBS $s$ at time $t$ \\
$\rbQueueEvo_s(t)$ & Resource queue at SBS $s$ at time $t$ \\
$\propCachePolicy_{f,s}(t)$ & Indicator vector of content $f$ at SBS $s$ at time $t$ \\
$\contentClassesSet_s$,$\ \noOfContentClasses_s$ & Set, Number of content classes at SBS \\
$\boldsymbol{M}_{ff'}^{i}$ & Similarity between content $f$ and $f'$ at $i$ \\
$\rbsSize$ & Size of rb set \\
$\rbsSet_s(t)$ & Set of RBs at SBS $s$ at time $t$ \\
$\numCacheUp$ & Number of cache updates \\
$\ueRequests_u(t)$ & Content requested by user $u$ at time $t$ \\
$\dataQueue_{ji}^f$ & Data queue of $i$ for content $f$ at $j$ \\
$\rate_{su}^{m,f}$ & Data rate of user $u$ by SBS $s$ over RB $m$  \\
$\channel_{su}^m$ & Channel between user $u$ by SBS $s$ over RB $m$ \\
\hline
\end{tabular}

\end{minipage} \hfill
\begin{minipage}{0.5\textwidth}
\begin{tabular}{|p{1.0cm} p{6.40cm}|}
\hline
\textbf{Symbols} & \textbf{Description}  \\
\hline
$\boldsymbol{\regret}_s$,$\ \boldsymbol{\regret}_c$ & Estimated regret vector at SBS $s$, cloud \\
$\cost_{ji}^f(t)$ & Transmission cost of $j$ to serve $i$ for content $f$ at time $t$ \\
$\qosReq_{s\ueRequests_{uf}}$ & QoS requirement vector of content $f$ at SBS $s$ requested by user $u$ \\
$\sbsClustTime$ & SBSs clustering time \\
$\cacheUpdateTime$ & Cache update time \\
$\uePrefTime$ & User preference time over contents \\
$\utility_s$,$\ \utility_c$ & Utility of SBS $s$, cloud \\
$\transBits_{ji}^f(t)$ & Transmitted bits of content $f$ requested by $i$ by $j$ at time $t$. \\
$\frthaulCap$ & Total fronthaul capacity \\
$\schedVar_{su}^f(t)$ & Scheduling decision of SBS $s$ by user $u$ at time $t$ \\ 
$\sysStateBig$,$\ \sysStateSet$ & System state, System state set \\
$\sbsTransDur$ & Duration of SBS transmission \\
$\locGlobTradeof$ & Local/global tradeoff of caching strategy of SBS and cloud \\
$\boldsymbol{\frthaulPerUe}_s$ & Fronthaul capacity vector of SBS $s$ \\ 
$\aggCost_s$,$\ \aggCost_c$ & Cost of SBS $s$, cloud \\
$\ImpPopSim$ & Impact of popularity on similarity \\
$\boldsymbol{\learningPop}_s$,$\ \boldsymbol{\learningPop}_c$ & Caching strategy of SBS $s$, cloud \\
$\cacheUpTime$ & Duration for cache update \\
$\CacheTempCoeff_s$ & Tradeoff parameter between exploration and exploitation \\
$\learParam_i^{\mathrm{s}}$ & Learning rates of SBS $s$ $i \in \{1,2,3\}$ \\
\hline
\end{tabular}
\end{minipage}
\end{table}
\vspace{-0.5cm}
\section{System Model}
\begin{figure}[t]%
\centering
\vspace{-1.2cm}
\includegraphics[scale = 0.525]{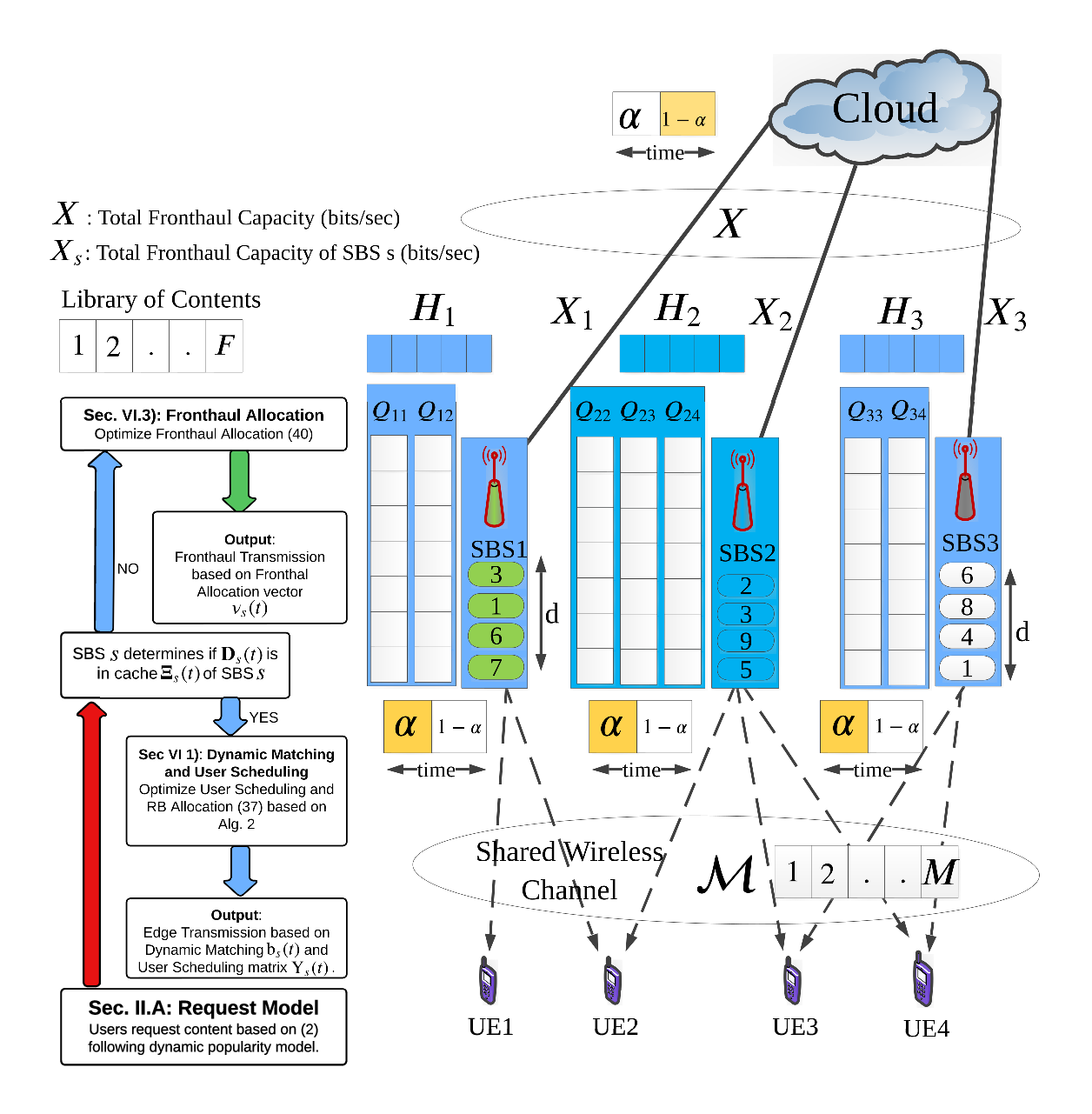}
\vspace{-0.7cm}
\caption{System Model.}
\vspace{-0.3cm}
\label{fig:Sys_Model}
\end{figure}

\vspace{-0.1cm}
Consider the downlink transmission of an ultra dense small cell network comprising of randomly deployed SBSs, $\sbsSet = \{1, ..., \sbsSize\}$ with intensity $\sbsDensity$ (SBSs per square meter) in $\mathbbm{R}^2$. Similarly, a set of user equipment(s) (UEs), $\ueSet = \{1, ..., \ueSize\}$, are also deployed randomly with intensity $\ueDensity$. The available system bandwidth is divided into $\rbsSize$ Resource Blocks (RBs) to serve UEs' requests as shown in Fig. \ref{fig:Sys_Model}. Hereafter, $\rbsSet = \{1, ..., \rbsSize\}$ represents the set of RBs shared among SBSs. \\
\indent Each SBS is equipped with a cache of size $\cacheSize$ where it stores contents from a content catalog $\contCatalogSet = \{1, ..., \contCatalog\}$ as shown in Fig. \ref{fig:Sys_Model}. We assume that the SBSs are not aware of the size of content catalog. Let $\frac {1} {\packetSize}$ (packets) be the average size of all the contents. Each content requires a certain QoS which depends on content type e.g., file, audio, video. Let $\boldsymbol{\insCache} = [\boldsymbol{\insCache}_s(t)]_{s \in \sbsSet}$ represent the SBS caches at time $t$ where $\boldsymbol{\insCache}_s(t) \subseteq \contCatalogSet$ represents the vector of cached contents by SBS $s$ at time $t$. Each SBS ranks contents based on the user's requests and utility termed as local content popularity. \\
\indent Each SBS is connected to the central controller/cloud via fronthaul links of fixed capacity $\frthaulCap$ through which SBSs update their cache. Let $\boldsymbol{\mathrm{\schedVar}}_s (t) = [\boldsymbol{\schedVar}_{su}(t)]_{u \in \ueSet}$ represent the scheduling matrix of SBS $s$ where $\boldsymbol{\schedVar}_{su}(t) = [\schedVar_{su}^f(t)]_{f \in \mathcal{F}}$ represent the scheduling vector of SBS $s$ for requests of user $u$ such that $\schedVar_{su}^f = 1$ indicates that the content $f$ requested by user $u$ is served by SBS $s$  at time $t$. The instantaneous data rate to serve request of user $u$ for content $f$ by SBS $s$ over RB $m$ is:
\begin{equation}
\label{eq:basicRate}
\textstyle \rate_{su}^{(m,f)}(t) = \textstyle \sbsTransDur \bwPerRB \schedVar_{su}^f(t) \log_2 \Big( \textstyle 1 + \frac { \textstyle \matchPolicy_{su}^m(t) \txPower_s^{m}(t) \channel_{su}^m(t) } { \textstyle \noise + \sum_{s' \in \sbsSet \setminus s} \matchPolicy_{s'u}^m(t) \txPower_{s'}^{m}(t) \channel_{s'u}^{m}(t) } \Big), 
\end{equation}
where $\noise$ represents the power spectral density of noise, $\channel_{su}^{m}(t)$ denotes the channel gain and $\txPower_s^{m}(t)$  denotes the transmit power between user $u$ and SBS $s$ over RB $m$ at time $t$, $\bwPerRB$ denotes the bandwidth per RB and $0 \leq \sbsTransDur \leq 1$ is the duration of SBS transmission\footnote{Note that $1 - \sbsTransDur$ is the duration of transmission on the fronthaul links}, respectively. The duration of the time slot is assumed to be long enough to complete the current transmission. \\
\indent Due to the constrained cache size, SBSs serve instantaneous user requests by retrieving the contents through the fronthaul links if the content is not cached by an SBS in the coverage of the user. Let $\frthaulPerUe_{s}(t)$ represent the fraction of fronthaul capacity allocated to SBS $s$ at time $t$ such that $\frthaulPerUe_s(t) \in [0,1]$. The cloud classifies contents based on the frequency of the received requests referred to as global content popularity. Further, the fronthaul links are used for transmitting the local content popularity (by SBSs) and global content popularity (to SBSs). The content requested by the $u$-th user at time $t$ is denoted by $\ueRequests_u(t) \in \contCatalogSet \cup \{0\}$ where $\ueRequests_u(t) = 0$ denotes no user request at time $t$, and $\matchPolicy_{su}^m(t)$ denotes the matching decision at SBS $s$ at time $t$ defined as:
\vspace{-0.2cm}
\theoremstyle{definition}
\begin{definition}{\textit{(Matching between users and RBs)}}
A matching matrix, denoted by $\boldsymbol{\mathrm{\matchPolicy}}_s = [\boldsymbol{\matchPolicy}_{su}(t)]_{u \in \ueSet}$, at an SBS $s$ defines the associated users and allocated resources of the SBS $s$. Here, $\boldsymbol{\matchPolicy}_{su}(t) = [\matchPolicy_{su}^{m}(t)]$ defines the association between user $u$ and SBS $s$ such that $\matchPolicy_{su}^{m}(t) = 1$ denotes RB $m$ allocated by SBS $s$ to user $u$ at time $t$ and $b = 0$ otherwise.  
\end{definition}
\vspace{-0.9cm}
\subsection{Spatio-temporal user request Model}
\indent Each user requests contents from a content library following the dynamic popularity model i.e., spatio-temporal model similar to Poisson Shot Noise Model (SNM) for simulations \cite{Ref28_Intro}. For this reason, we assume that each user generates content requests following the Zipf distribution i.e., $\boldsymbol{\ueReqDist}_u = [\mathbbm{\ueRequests}_{uf}]_{f \in \contCatalogSet}$ where the probability of requesting content $f$ is:
\begin{equation}
\textstyle \mathbbm{\ueRequests}_{uf_u} = \frac {\textstyle 1} {\textstyle f_{u}^{\pwrLawExp_u}} \Big( \textstyle \sum_{i=1}^{\contCatalog} \frac {\textstyle 1} {\textstyle i^{\pwrLawExp_u}} \Big)^{-1},
\end{equation}
where $\pwrLawExp_u$ is the power law exponent and $f_{u} = E_{u}(f)$ is a permutation/ordering function defined for user $u$ \cite{Ref1:Pop_Sec}. For creating request distribution variation in space, the power law exponent is assumed to have a different value for each user. In addition, each user has its own content preference i.e., a content highly preferred by a user may not have the same preference for the other user(s). On the contrary, to vary the request distribution in time, the user preference remains fixed for the duration $\uePrefTime$. After $\uePrefTime$, the user alters his preference by changing the power law exponent i.e., $\pwrLawExp_u(t+1) = \pwrLawExp_u(t) \pm \PopPowerLawExp_u^2$ such that $\pwrLawExp_u(t)>0, \forall t$ where $\PopPowerLawExp_u^2$ denotes the variance of the power law exponent and defines the change in user interests \cite{Ref18_CollaborativeCacheEdge3}. The content demand vector at SBS $s$ is $\boldsymbol{\contDemand}_s = [\contDemand_{sf}(t)]_{f \in \contCatalogSet}$ such that $\contDemand_{sf}(t) = \sum_{u \in \ueCoverage_s(t)} \mathbb{1}_{\{\ueRequests(t) = f\}}$ where $\mathbb{1}_{x}$ is the indicator function and $\ueCoverage_s(t)$ represents UEs in the coverage of SBS $s$ at time $t$. Accordingly, the content demand matrix is denoted by $\boldsymbol{\mathrm{\contDemand}} = [\boldsymbol{\contDemand}_s(t)]_{s \in \sbsSet}$.
\vspace{-0.7cm}
\subsection{Request and Resource Queues}
\label{ssec: Task_Resource}
Due to the shared RBs among SBSs and mutual interference, an SBS may not immediately fulfill the user request. Likewise, the cloud does not always serve the user request immediately due to the constraint fronthaul capacity. As a result, the dynamics of the request queue at the SBS and cloud are:
\vspace{-0.2cm}
\begin{align}
\textstyle \dataQueue_{su}^f(t+1) & = \textstyle \mathrm{max}[\dataQueue_{su}^f(t) - \textstyle \sum_{m=1}^{\rbsSize} \rate_{su}^{(m,f)}(t), 0] + \textstyle \packetArrRate_{su}^f(t) \ \forall s, u, f, \label{eq: queue_eq1} \\
\textstyle \dataQueue_{cs}^f(t+1) & = \textstyle \mathrm{max}[\dataQueue_{cs}^f(t) - \textstyle (1-\sbsTransDur) \frthaulPerUe_{s}^f \frthaulCap , 0] + \textstyle \frac {\textstyle 1} {\textstyle \packetSize_f} \sum_{u=1}^{\ueSize} \mathbbm{1}_{\{q_u(t) \not\in \insCache_s(t) \} } \ \forall s, f, \label{eq: queue_eq3}
\end{align}
where $\dataQueue_{su}^f$, $\dataQueue_{cs}^f$ denotes the number of packets of file $f$ destined for user $u$, SBS $s$ at time $t$, $\packetArrRate_{su}^f(t)$, $\packetArrRate_{cs}^f(t)$ denotes the number of packets of content $f$ for the requested content of user $u$ arriving at SBS $s$, cloud, $\frac {\textstyle 1} {\textstyle \packetSize_f}$ denotes the size of content $f$, $\mathbbm{1}_{\{q_u(t) \not\in \insCache_s(t) \}}$ denotes the requested content not in cache of SBS $s$ and $\frthaulPerUe_s^f$ denotes the fraction of fronthaul capacity for content $f$ for SBS $s$ such that $\sum_{f=1}^{\contCatalog} \sum_{s=1}^{\sbsSize} \frthaulPerUe_s^f = 1$. In addition, let $\boldsymbol{\packetArrRate}_{su}(t) = [\packetArrRate_{su}^f(t)]_{f \in \contCatalogSet}$ represents the vector of packet arrival at SBS $s$ and $\boldsymbol{\dataQueue}_{su}(t) = [\dataQueue_{su}^f(t)]_{f \in \contCatalogSet}$ denote the task queue vector at SBS $s$. Let $\packetArrRate_{\mathrm{max}}$ be the maximum number of new packets arriving at any SBS $s \in \sbsSet$ which depends on $\boldsymbol{\insCache}_s(t)$ (in case the content is not cached by any SBS in the coverage of UE) and the QoS requirement of the requested content. If the content is cached by an SBS, $\packetArrRate_{su}^f(t)$ is equal to the number of packets associated with the requested content. \\
\indent When an SBS fulfills a user request, the RBs become available for all SBSs. Let $\rbsArrRate_s(t) \in [0, \rbsSize]$ be the arrival rate of RBs for SBS $s$ at time $t$ and $\rbQueueEvo_s(t)$ denotes the number of RBs held by SBS $s$ at time $t$. Hence, the dynamics of resource queue is:  
\vspace{-0.2cm}
\begin{equation}
\label{eq: queue_eq2}
\textstyle \rbQueueEvo_{s}(t+1) = \textstyle \rbQueueEvo_s(t) - \textstyle \sum_{u \in \ueCoverage_s(t)} |\boldsymbol{\matchPolicy}_{su}(t)| + \textstyle \rbsArrRate_s(t) \ \forall s.
\vspace{-0.2cm}
\end{equation}
Based on the matching policy, the latency in terms of time at SBS $s$ to serve $u$ for content $f$ is defined as:
\begin{equation}
\vspace{-0.2cm}
\label{eq:Reward_eq1}
\textstyle \cost_{su}^f(\boldsymbol{\mathrm{\sysStateBig}}(t), \boldsymbol{\insCache}_s(t), \boldsymbol{\mathrm{\matchPolicy}}(t)) = \textstyle \frac {\textstyle \numBitsPerPacket \transBits_{su}^f(t)} {\textstyle \sbsTransDur \bwPerRB \schedVar_{su}^f(t) \mathlarger{\sum}_{m =1}^{\rbsSize} \log_2 \Big( 1 + \frac { \textstyle {\matchPolicy_{su}^m(t) \txPower_s^{m}(t) \channel_{su}^{m}(t)} } { \textstyle \noise + \sum_{s' \in \sbsSet \setminus s} \matchPolicy_{s'u}^m(t) \txPower_{s'}^{m}(t) \channel_{s'u}^{m}(t) } \Big) + \textstyle c }, 
\end{equation}
where $\numBitsPerPacket$ is a constant representing the number of bits per packet, $\transBits_{su}^f(t) \in [0, \dataQueue_{su}^f(t)]$, $\boldsymbol{\mathrm{\sysStateBig}}(t) = [\sysStateSmall_s(t), \sysStateSmall_{-s}(t)]$ represents the time-varying channel conditions called system state such that $\sysStateSmall_{s}(t) = (\boldsymbol{\packetArrRate}_s(t), \rbsArrRate_s(t), \boldsymbol{\channel}_s(t)) \in \sysStateSet_s$ is an i.i.d process of $I$ finite value $\sysStateSet_s = \{\sysStateSmall_1, ..., \sysStateSmall_I\}$, $\sysStateSmall_{-s}(t)$ represents the system state of SBSs excluding SBS s \footnote{Hereafter subscript $-i$ represents all SBSs except the $i$th SBS.} and $\contDelay$ is a constant to control the upper bound of the cost. For a content that is not cached by SBSs, the cost in terms of time at the cloud to to serve SBS $s$ for content $f$ is:
\vspace{-0.2cm}
\begin{equation}
\textstyle \cost_{cs}^f(\boldsymbol{\insCache}_s(t), \boldsymbol{\frthaulPerUe}(t)) = \frac {\textstyle \numBitsPerPacket \transBits_{cs}^f(t)} {\textstyle (1 - \sbsTransDur) \frthaulPerUe_s^f(t) \frthaulCap + c} ,
\vspace{-0.2cm}
\end{equation}
where $\transBits_{cs}^f(t) \in [0, \dataQueue_{cs}^f(t) + \frac{\textstyle 1} {\textstyle \packetSize_f}]$ and $\frthaulPerUe_{s}^f$ denotes the fraction of fronthaul capacity allocated for $\dataQueue_{cs}^f(t)$ such that $\sum_{s \in \sbsSet} \sum_{f \in \contCatalogSet} \frthaulPerUe_s^f(t) = 1$.
\vspace{-0.5cm}
\section{Queuing Latency Minimization Problem}
\label{sec: Prob_Form}
The performance of the caching strategy is measured by the availability of requested content in the cache of SBSs and the time to transmit the requested content. Our objective is to determine a joint caching policy and resource allocation per SBS which minimize the queuing latency in serving UEs' requests while ensuring UEs' QoS requirement. At each time $t$, SBSs allocate RBs to serve users' requests depending on dynamic channel conditions and existence of content in SBS(s) cache. This decision is defined by a matching between the request of user $u$ and SBS $s$ over RB $m$. As a result, the queuing latency (defined as the time to transmit the content) depends on the caching strategy, channel conditions as well as QSI of SBS. On the contrary, UE requests for contents that are not cached by the SBSs will be fetched from the cloud with an additional cost in terms of service delay. Therefore, the instantaneous cost of SBS depends on both caching policy and resource allocation. \\
\indent Depending on the QoS (latency) requirement, an SBS may allocate single or multiple RBs to the user queues. However, an RB can be allocated to a single user queue at an SBS. Therefore, each SBS implements \textit{one-to-many matching} between user queues and RBs. When $\sysStateSmall_s(t) = \sysStateSmall_i$, the matching matrix $\boldsymbol{\mathrm{\matchPolicy}}_s(t) \subset \rbsSet_s(t)$ where $\rbsSet_s(t)$ denotes the RBs used by SBS $s$ at time $t$ such that $|\rbsSet_s(t)| = \rbQueueEvo_s(t)$. Let $\boldsymbol{\matchPolicy}_s^{\mathrm{max}}(t) \overset{\Delta}{=} \max_{\boldsymbol{\mathrm{\matchPolicy}}_s(t) \in \rbsSet_s(t)} \lVert \rbsSet_s(t) \rVert_{\infty} \leq \rbQueueEvo_s(t)$ represents the maximum number of RBs assigned to any user queue at any time. Based on the system state, caching and matching policy $\boldsymbol{\mathrm{\matchPolicy}}(t) = [\boldsymbol{\mathrm{\matchPolicy}}_s(t), \boldsymbol{\mathrm{\matchPolicy}}_{-s}(t)]$ such that $\boldsymbol{\mathrm{\matchPolicy}}_{-s}(t)$ represents the matching of other SBSs except SBS $s$, the instantaneous rate matrix $\boldsymbol{\mathrm{\rate_{su}}}(\boldsymbol{\mathrm{\sysStateBig}}(t), \boldsymbol{\insCache}_s(t), \boldsymbol{\mathrm{\matchPolicy}}(t))$ has the following properties:
\begin{itemize}
\item $\boldsymbol{\mathrm{\rate_{su}}}(\boldsymbol{\mathrm{\sysStateBig}}(t), \boldsymbol{\insCache}_s(t), \boldsymbol{\mathrm{\matchPolicy}}(t)) \in [0, \rate_{\mathrm{max}}]$ where $\rate_{\mathrm{max}}$ represents the rate under no interference.
\item If $\rbQueueEvo_s(t) = 0$, no matching is possible i.e., $\boldsymbol{\mathrm{\matchPolicy}}_s(t) = \boldsymbol{\mathrm{0}}$. Therefore, $\boldsymbol{\mathrm{\rate_{su}}}(\boldsymbol{\mathrm{\sysStateBig}}(t), \boldsymbol{\insCache}_s(t), \boldsymbol{\mathrm{0}}(t)) = 0$.
\end{itemize}

\indent From \eqref{eq:Reward_eq1}, it can be noted the incurred cost is a function of caching policy and matching. If the content has been cached by an SBS, the incurred cost depends on the matching due to the shared radio resources among SBSs. On the other hand, the cost increases if no SBS in the coverage of the user caches the requested content. As a result, the content is retrieved from the cloud which increases the cost even more due to shared fronthaul capacity among SBSs. Therefore, caching and matching problems are coupled together. Solving both problems simultaneously is challenging due to the stochastic nature of user requests, QSI and wireless channel. Hence, we leverage two time scales to the caching and matching policy: In the first phase, the objective is to design a caching strategy based on time-varying matching policy which minimizes the latency corresponding to the UEs' requests at SBSs. In the second phase, each SBS implements a matching policy based on the fixed caching strategy.
\subsubsection{Caching Strategy}
\label{sssec:Caching_Strategy}
Based on the user demand, each SBS implements a caching policy that aims to maximize the instantaneous cache hits. The caching policy determines the contents to cache based on time-varying content popularity. At every time $t$, each SBS receives requests from UEs. Based on these requests, the SBSs learn content popularity over time and update their cache accordingly. For a small number of requests, the popularity at SBSs may not be determined accurately. Hence, it is important to have enough statistics for an efficient caching policy. Simultaneously, the problem becomes more challenging when the number of most popular contents are larger than cache size. Further, determining time-varying content popularity becomes computationally expensive for a large number of contents and user requests. To overcome this issue, we assume that the caching policy may be devised by the central controller which determines content popularity for all the SBSs. In addition, the SBSs also groups contents with similar popularity into different clusters/classes and learn class popularity rather than content popularity. Let $\contentClassesSet_s = \{1, ..., \noOfContentClasses_s\}$ be the set of classes at SBS $s$ where $\noOfContentClasses_s$ denotes the number of classes at SBS $s$. The contents may be classified based on global popularity, local popularity or a weighted combination of both variants. However, classifying contents based on the global popularity incurs additional cost as each SBS transmits and receive data from central controller as well the demand vector to a central controller. It is worth mentioning that classifying contents based on global popularity yields homogeneous classes across the whole network i.e., $\noOfContentClasses_s = \noOfContentClasses_{s'}$, $\forall s, s' \in \sbsSet$. On the contrary, the caching policy needs to be revised if the instantaneous hit rate corresponding to the current caching policy is low. However, varying the caching policy frequently incurs high cost in terms of fronthaul access. Therefore, the caching policy remains fixed for a finite time period $T_2$. Let $\propCachePolicy_{f,s}(\boldsymbol{\learningPop}_s(t)) = 1$ with probability $\boldsymbol{\learningPop}_s(t)$ indicates the content $f$ needs to be cached at SBS $s$ at time $t$ and $\propCachePolicy_{f,s}(\boldsymbol{\learningPop}_s(t)) = 0$ otherwise where $\boldsymbol{\learningPop}_s(t) = [\learningPop_{sf}(t)]_{f \in \contCatalogSet}$ represents the caching strategy of SBS $s$ at time $t$. With the fronthaul capacity of $\frthaulCap_s(t)$ for SBS $s$ and assuming $\numCacheUp$ cache updates, the time required for cache update at SBS $s$ is:
\begin{equation}
\vspace{-0.1cm}
\textstyle \cacheUpTime_s( \insCache_s(t), t) = \textstyle \begin{cases} & \textstyle \numBitsPerPacket \sum_{f=1}^{\contCatalog} \frac {\textstyle \propCachePolicy_{f,s}(\boldsymbol{\learningPop}_s(t)) (1 - \mathbb{1}_{f \in \boldsymbol{\insCache}_s(t)})} {\textstyle \frac {\textstyle 1} {\textstyle \packetSize_f} \textstyle \frthaulCap_s(t)} \ \mathrm{with} \ \ \sum_{f = 1}^{\contCatalog} \textstyle \propCachePolicy_{f,s}(\boldsymbol{\learningPop}_s(t)) \leq \numCacheUp, \ \ t = n \cacheUpdateTime, \\
\vspace{-0.1cm}
& \textstyle 0 \ \ \ \ \ \ \ \ \ \ \ \ \ \ \ \ \ \ \ \ \ \ \ \ \mathrm{otherwise}, 
\end{cases}
\vspace{-0.2cm}
\end{equation}
\vspace{-0.1cm}
where $n\cacheUpdateTime$ represents the instances of cache update with $n \in \mathbbm{Z}$.
\subsubsection{Matching Policy} 
\begin{figure}[t]%
\centering
\captionsetup{justification=centering}
\vspace{-1.0cm}
\includegraphics[scale = 0.45]{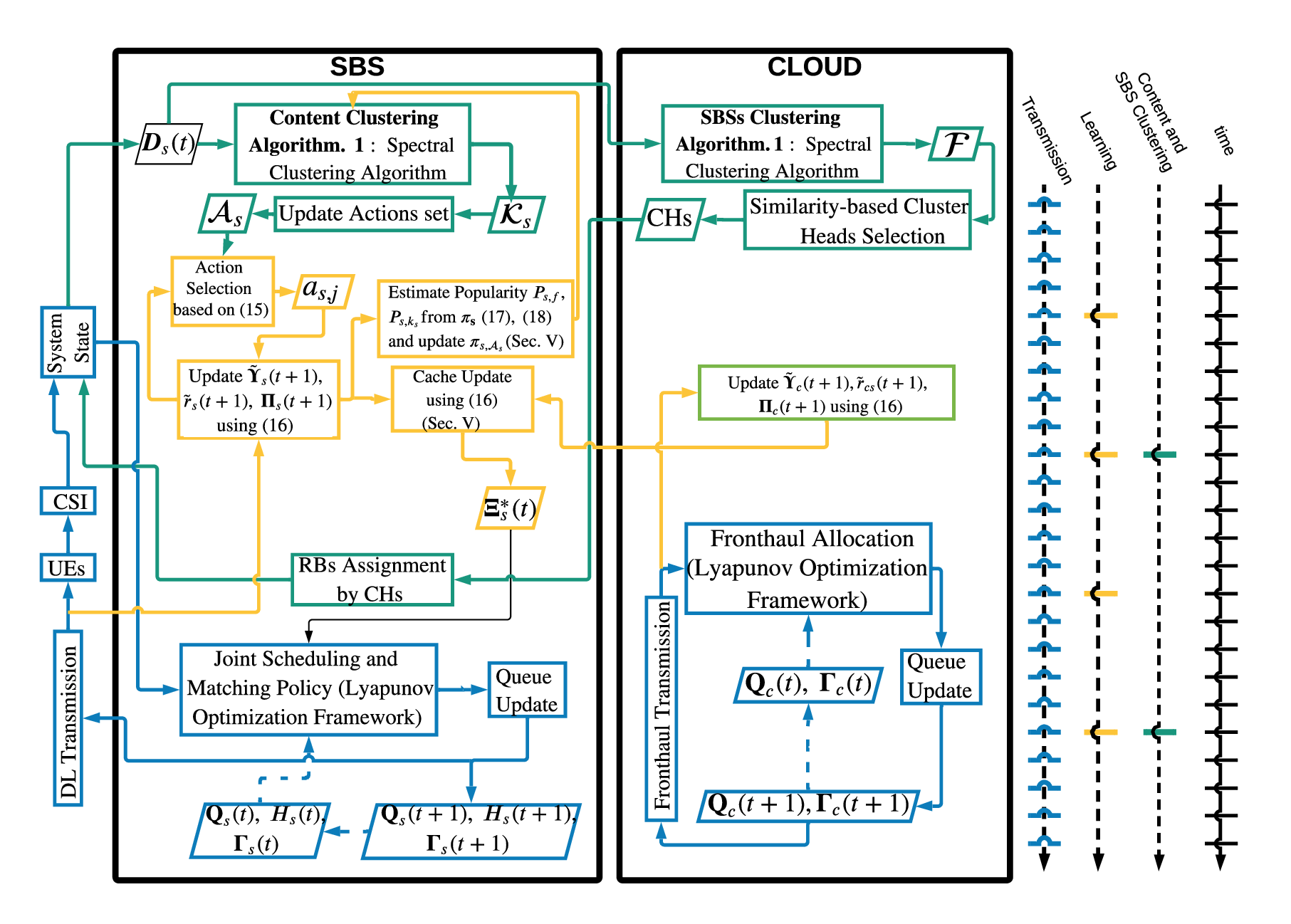}
\vspace{-0.6cm}
\caption{The proposed Joint Reinforcement Learning caching and Edge Transmission algorithm including the interaction between the SBSs and Cloud}
\vspace{-0.2cm}
\label{fig:Flow}
\end{figure}

In order to implement a matching policy, a user-SBS assocation is required for which caching policy and QSI are considered. We assume that the user associates to the SBS with cached content. For a requested content cached by multiple SBSs, user associates to the nearest SBS with minimum queue length. We let $s(u)$ be the SBS $s$ that serves user $u$. In the rest of the paper, the SBS to which the UE associates is termed \textit{anchor SBS}. We assume that the user remains associated to the anchor SBS as long as the packets associated to the previously requested content are not fully transmitted. Once the users are associated, the anchor SBS needs to allocate resources to the associated users while ensuring desired QoS. For this purpose, each SBS implements a matching policy to serve user request where it matches user queues to the resource queues/set of RBs. At every time $t$, each SBS first performs random matching on the available RBs. Due to the absence of coordination mechanism among SBSs, each SBS experiences excessive interference from the neighboring SBSs especially in ultra dense environments. To overcome this issue, a running time average state based learning mechanism is used which minimizes the cost expressed in \eqref{eq:Reward_eq1}. \\
\indent For computation efficient caching strategy for a large number of contents and mitigate interference, the central controller groups SBSs with similiar content popularity and distance dissimilarity into different clusters. Let $\sbsClusterSet = \{\sbsClusterSet_1, ..., \sbsClusterSet_{\sbsSize}\}$ represents the set of SBS clusters where intra-cluster coordination mechanism exists between SBSs. \\
\indent By leveraging different time-scales, we assume that content clustering is much slower than the cache update procedure. Hence, for each SBS $s \in \sbsSet$, clustering remains fixed for a period $\sbsClustTime$ while the caching policy remains fixed for a time period of $\cacheUpdateTime$ where $\sbsClustTime > \cacheUpdateTime$, and determined based on learning demand profile for the content library as shown in Fig. \ref{fig:Flow}. \\
\indent When the SBS classifies contents based on global popularity, it transmits the local demand vector to the central controller which incurs additional delay. During the local demand vector transmission, no user scheduling is performed at the global content clustering time instant i.e., $\sbsClustTime, 2\sbsClustTime, ...$. As no user scheduling is performed, no cost is incurred at the global clustering time instant i.e., $\boldsymbol{\cost}_s(n\sbsClustTime) = 0, n \in \mathbb{Z^{+}}$.
For each cluster of SBSs $\sbsClusterSet_i \in \sbsClusterSet$, the average per-cluster cost for caching policies $\boldsymbol{\insCache}_s^{\sbsClustTime} = [\boldsymbol{\insCache}_s(0), \boldsymbol{\insCache}_s(\sbsClustTime), \boldsymbol{\insCache}_s(2\sbsClustTime), ..., \boldsymbol{\insCache}_s(n\sbsClustTime)]$ with $n\sbsClustTime < \cacheUpdateTime$ and $n \in \mathbbm{Z}$, matching policies $\boldsymbol{\mathrm{\matchPolicy}}_s = [\boldsymbol{\mathrm{\matchPolicy}}_s(0), \boldsymbol{\mathrm{\matchPolicy}}_s(1), ..., \boldsymbol{\mathrm{\matchPolicy}}_s(\cacheUpdateTime-1)]$ and fronthaul allocation $\boldsymbol{\frthaulPerUe}_s = [\boldsymbol{\frthaulPerUe}_s(0), \boldsymbol{\frthaulPerUe}_s(1), ..., \boldsymbol{\frthaulPerUe}_s(\cacheUpdateTime-1)]$ is the aggregate cost of the associated UEs in the cluster i.e.,
\vspace{-0.2cm}
\begin{equation}
\label{eq:Cluster_Utility}
\textstyle \bar{\aggCost}_{\sbsClusterSet_i}(\boldsymbol{\insCache}_s^{\sbsClustTime}, \boldsymbol{\mathrm{\matchPolicy}}_s^{\cacheUpdateTime}) = \textstyle \frac {\textstyle 1} {\textstyle | \sbsSet_{\sbsClusterSet_i} |} \sum_{\forall s \in \sbsClusterSet_i} \textstyle \bar{\aggCost}_{s}(\boldsymbol{\insCache}_s^{\sbsClustTime}, \boldsymbol{\mathrm{\matchPolicy}}_s^{\cacheUpdateTime}) + \textstyle \cacheUpTime_s, \ \ \ \ \textstyle \sbsSet_{\sbsClusterSet_i} = \{s: \forall s \in \sbsClusterSet_i\},  
\vspace{-0.2cm}
\end{equation}
where 
\vspace{-0.1cm}
\begin{equation}
\label{eq:sbss_utility}
\textstyle \bar{\aggCost}_{s}(\boldsymbol{\insCache}_s^{\sbsClustTime}, \boldsymbol{\mathrm{\matchPolicy}}_s^{\cacheUpdateTime}) = \textstyle \frac {\textstyle1} {\textstyle \sbsClustTime} \textstyle \sum_{\tau_1 = 0}^{\sbsClustTime-1} \frac {\textstyle 1} {\textstyle \cacheUpdateTime} \textstyle \sum_{\tau_2 = 0}^{\cacheUpdateTime-1} \textstyle \lVert \boldsymbol{\mathrm{\cost}}_{s}(\boldsymbol{\mathrm{\sysStateBig}}(\tau_2), \boldsymbol{\insCache}_s(\tau_1), \boldsymbol{\mathrm{\matchPolicy}}(\tau_2)) \rVert_{\infty} + \textstyle \frac {\textstyle 1} {\textstyle \sbsClustTime} \textstyle \sum_{\tau_1 = 0}^{\sbsClustTime-1} \textstyle \lVert \boldsymbol{\mathrm{\dataQueue_s}}(\tau_1) \rVert_{\infty},
\end{equation}
\vspace{-0.1cm}
with $\boldsymbol{\mathrm{\dataQueue_s}}(t) = [\boldsymbol{\dataQueue_{su}(t)}]_{u \in \ueSet}$ and $\lVert . \rVert$ represents the maximum norm. Further, the cost at the cloud is given as: 
\begin{equation}
\label{eq:cloud_utility}
\textstyle \bar{\aggCost}_{c}(\boldsymbol{\insCache}_s^{\sbsClustTime}, \boldsymbol{\frthaulPerUe}^{T_2}) = \textstyle \frac {\textstyle 1} {\textstyle \sbsClustTime} \textstyle \sum_{\tau_1 = 0}^{\sbsClustTime-1} \textstyle \frac {\textstyle 1} {\textstyle \cacheUpdateTime} \textstyle \sum_{\tau_2 = 0}^{\cacheUpdateTime-1}  \textstyle \lVert \big( \boldsymbol{\mathrm{\cost}}_{c}( \boldsymbol{\insCache}_s(\tau_1), \frthaulPerUe) \rVert_{\infty} + \textstyle \frac {\textstyle 1} {\textstyle \sbsClustTime} \textstyle \sum_{\tau_1 = 0}^{\sbsClustTime-1} \textstyle \lVert \boldsymbol{\mathrm{\dataQueue_c}}(\tau_1) \rVert_{\infty},
\vspace{-0.1cm}
\end{equation}
where $\boldsymbol{\mathrm{\cost}}_{c} = [\boldsymbol{J}_{cs}]_{s \in \sbsSet}$ such that $\boldsymbol{J}_{cs} = [J_{cs}^f]_{f \in \contCatalogSet}$ and $\boldsymbol{\mathrm{\cost}}_{s} = [\boldsymbol{J}_{su}]_{u \in \ueSet}$ such that $\boldsymbol{J}_{su} = [J_{su}^f]_{f \in \contCatalogSet}$. The main aim is to minimize the average total delay over the fronthaul and access links i.e., $\bar{\aggCost}_{\mathrm{T}}(\boldsymbol{\insCache}_s^{\sbsClustTime}, \boldsymbol{\mathrm{\matchPolicy}}_s^{\cacheUpdateTime}, \boldsymbol{\frthaulPerUe}^{T_2}) = \bar{\aggCost}_{c}(\boldsymbol{\insCache}_s^{\sbsClustTime}, \boldsymbol{\frthaulPerUe}^{T_2}) + \frac {1} {|\sbsClusterSet|} \sum_{\forall \sbsClusterSet_i \in \sbsClusterSet}  \bar{\aggCost}_{\sbsClusterSet_i}(\boldsymbol{\insCache}_s^{\sbsClustTime}, \boldsymbol{\mathrm{\matchPolicy}}_s^{\cacheUpdateTime})$ where $|\sbsClusterSet|$ represents the size of $\mathcal{C}$ \footnote{Subsequently, $|.|$ represents the cadinality.}. The expectation over clusters is taken due to simultaneous transmission by all clusters. Hence, the cost minimization problem is:
\vspace{-1.2cm}

\begin{subequations}
\label{eq: Obj_Func}
\centering
\begin{align}
    \underset{\textstyle \boldsymbol{\insCache}^{\sbsClustTime}, \mathrm{\boldsymbol{\matchPolicy}}^{\cacheUpdateTime},\boldsymbol{\schedVar^{\cacheUpdateTime}}, \boldsymbol{\frthaulPerUe}^{\cacheUpdateTime}} {\text{minimize}} \ \ \ & \textstyle \bar{\aggCost}_{\mathrm{T}}(\boldsymbol{\insCache}_s^{\sbsClustTime}, \boldsymbol{\mathrm{\matchPolicy}}_s^{\cacheUpdateTime}, \boldsymbol{\frthaulPerUe}^{T_2}), \label{eq:Obj_Func1} \\
    \text{subject to} \ \ & \textstyle |\boldsymbol{\insCache}_s(t)| \leq \textstyle \cacheSize, \ \forall s \in \sbsSet, \label{eq:Obj_Func2} \\
		\ \ & \textstyle \mathbbm{P} \big\{\textstyle \frac {\textstyle \dataQueue_{su}^f(t)} {\textstyle \bar{\packetArrRate}_{su}^f} > \textstyle \qosReq_{sq_{uf}} \big\} \leq \textstyle , \ \ \forall u \in \ueSet, f \in \contCatalogSet, t, \label{eq:Obj_Func3} \\
		\ \ & \textstyle \mathbbm{P} \big\{ \textstyle \packetSize_f \dataQueue_{cs}^f(t) > \textstyle \qosReq_{sq_{uf}} \big\} \leq \textstyle \epsilon_s, \ \forall s \in \sbsSet, f \in \contCatalogSet, t, \label{eq:Obj_Func4} \\
		\ \ & \textstyle \sum_{u \in \ueCoverage_s} |\boldsymbol{\matchPolicy}_{su}(t)| \leq \textstyle \rbQueueEvo_s(t), \ \ \forall s \in \sbsSet, \label{eq:Obj_Func5} \\
		\ \ & \textstyle \matchPolicy_{su}^m(t) \leq \textstyle \sum_{f=1}^{\contCatalog} \textstyle \schedVar_{su}^f(t), \ \ \ \forall s \in \sbsSet, u \in \ueSet, m \in \rbsSet_s, t, \label{eq:Obj_Func6} \\
		\ \ & \textstyle \schedVar_{su}^f(t) \in \textstyle \{0,1\}, \ \ \forall u \in \ueSet, f \in \contCatalogSet, t, \label{eq:Obj_Func7} \\
		\ \ & \textstyle \frthaulPerUe_s^f \in \textstyle [0,1], \ \ \forall s \in \mathcal{S}, f \in \contCatalogSet, \label{eq:Obj_Func8} \\
		\ \ & \textstyle \mathrm{\boldsymbol{\matchPolicy}}_s(t) \in \textstyle \rbsSet_s(t), \ \ \forall s \in \sbsSet, \label{eq:Obj_Func9} \\	
		\ \ & \textstyle \eqref{eq: queue_eq1}, \eqref{eq: queue_eq3}, \eqref{eq: queue_eq2}, \label{eq:Obj_Func10} \ \forall t,
\end{align}
\end{subequations}
where \eqref{eq:Obj_Func2} indicates that the SBS caches content equal to the cache size, \eqref{eq:Obj_Func3} and \eqref{eq:Obj_Func4} denotes the latency and reliability constraint such that $\qosReq_{sq_{uf}}$ denotes the QoS requirement in terms of latency associated with content $f$ requested by user $u$ at SBS $s$, $\epsilon_u$ and $\epsilon_u$ represents the latency violation probability at $u$ and $s$, and $\bar{\packetArrRate}_{su}^f$ denotes the average packet arrival of file $f$ at SBS $s$ for user $u$,  \eqref{eq:Obj_Func5} indicates that an SBS cannot allocate radio resources more than the available radio resources and \eqref{eq:Obj_Func6} reveals that an SBS can schedule more contents on one RB. As the content popularity profile varies slowly over time compared to channel conditions, problem \eqref{eq: Obj_Func} is solved by separating it into two parts: \emph{i}) caching policy is obtained subject to \eqref{eq:Obj_Func2}, \emph{ii}) optimal matching using Lyapunov framework subject to \eqref{eq:Obj_Func5}-\eqref{eq:Obj_Func10}.
\LinesNumberedHidden{
\begin{algorithm}[t]
\footnotesize
\caption{Content Clustering and Cache Update}
\label{algo:algo1}
\DontPrintSemicolon 
\textbf{Input}: 
Observed $\boldsymbol{\learningPop}_s(t)$ and Global/local tradeoff parameter $\locGlobTradeof$. \\
\KwResult{Content cluster at SBSs $\contentClassesSet_s = \{1, ... \noOfContentClasses_s\}$, $\forall s \in \sbsSet$.}

\textbf{Algorithm:}

\textbf{Phase I - Spectral Clustering Algorithm;}
\begin{itemize}
		\item Compute the similarity matrix $\boldsymbol{\mathrm{M}}(t)$ based on \eqref{eq:Similarity} and perform spectral clustering algorithm as per \cite{Ref18_CollaborativeCacheEdge4}.
\end{itemize}

\textbf{Phase II - Regret Learning and Cache Update;}
\begin{itemize}
		\item Each SBS learns the probability distribution vector $\boldsymbol{\learningPop}_{s}$ based on \eqref{eq:Regret}.
    \item The cloud learns the probability distribution vector $\boldsymbol{\learningPop}_c$ based on \eqref{eq:Regret}.
		\item Each SBS updates its cache based on the mixed distribution $\boldsymbol{\learningPop}' = (1-\locGlobTradeof)\boldsymbol{\learningPop}_{s} + \locGlobTradeof \boldsymbol{\learningPop}_c$.
\end{itemize}
\end{algorithm}
}

\vspace{-0.5cm}
\section{Content Clustering and SBS Clustering}
\vspace{-0.1cm}
\subsection{Content Clustering}
In a real system, there exists a correlation among contents requests i.e., request of a content over a spatial location is nearly similar to the request of one or more contents in the same geographical area \cite{Ref14_Intro}. On the hand hand, caching decisions becomes extremely challenging when the library size becomes very large due to the increased computational complexity. This suggests grouping contents based on their demands as a solution to improve caching decisions. By observing the content demand over a finite time period, contents are clustered into different classes where contents in the same class have similar popularity. Thus, caching decision $\insCache_s(t)$ in \eqref{eq: Obj_Func} is solved over classes rather than contents. As a result, the caching decision becomes more robust and computationally less expensive compared to the unclustered method. In this work, a similarity measure between demand vectors is used to cluster contents into classes. Since, content similarity varies slowly over time, content clustering is a slower process than cache update. In other words, the content clustering remains fixed for a period $\sbsClustTime > \cacheUpdateTime$ where $\cacheUpdateTime$ is the cache update time. Towards this, a similarity measure $\boldsymbol{M}_{ff'}^{\mathrm{C}}(t)$ at time $t$ at SBS $s$ is defined as:
\begin{equation}
\label{eq:Similarity}
\textstyle \boldsymbol{M}_{ff'}^{s}(t) = \textstyle \exp \left(- \textstyle \frac {\textstyle |\contDemand_{sf}(t) \popularity_{sf}(t) - \textstyle \contDemand_{sf'}(t) \popularity_{sf'}(t)|^2} {\textstyle 2 \sigma_c^2} \right), \ \ \ \forall f' \in \contCatalogSet,
\end{equation}
where $\popularity_{sf}(t)$ is the popularity of file $f$ at SBS $s$ at time $t$ and $\sigma_c^2$ controls the impact of popularity on similarity and the similarity matrix is given by $\boldsymbol{\mathrm{M}^s}(t) = [\boldsymbol{M}_{ff'}^{s}(t)]_{f,f' \in \contCatalogSet}$. In this work, spectral clustering technique is used to perform content clustering \cite{Ref1:Clustering1} that exploits the probability distribution of content and the variance of the similarity matrix to form content classes. Due to the use of instantaneous content demand vector and popularity profile, the spectral clustering algorithm efficiently distributes contents into distinct clusters.
\vspace{-0.35cm}
\subsection{SBSs Clustering}
To serve instantaneous users' requests, the SBSs schedule users and allocates RBs through resource allocation. However, efficient resource allocation and instantaneous user scheduling requires perfect CSI, which may introduce unacceptable signaling overhead in ultra-dense networks. To minimize the effect of imperfect CSI and the additional signal overhead, the SBSs are classified into numerous clusters where SBSs within a cluster coordinate with one another to use orthogonal RBs. The cloud/central controller is responsible for clustering SBSs with similar characteristics i.e., SBSs with similar number of content requests are clustered together. To identify the similarities between SBSs, each SBS transmits the $\boldsymbol{\contDemand}_s(t)$ to the cloud. Thereafter, the cloud identifies the similarities between SBSs using cosine similarity measure as follows:
\begin{equation}
\textstyle \rbsSize_{ss'}^{\textrm{S}}(t) = \textstyle \frac {\textstyle \boldsymbol{\contDemand}_s(t) \cdot \boldsymbol{\contDemand}_{s'}(t)} {\textstyle \left\Vert \boldsymbol{\contDemand}_s(t) \right\Vert \left\Vert \boldsymbol{\contDemand}_{s'}(t) \right\Vert }.
\end{equation} 
\indent Based on the similarities, the cloud clusters the SBSs by employing spectral clustering algorithm. In addition, the cloud selects an SBS from the cluster (termed as \textit{cluster head}) which coordinates the utilization of radio resources as well as user scheduling within a cluster. The cloud selects cluster head based on the spectral clustering method described in Alg. \ref{algo:algo1} i.e., the SBS with the maximum cumulative similarity measure is selected as the cluster head. \\
\indent As mentioned in Sec. \ref{sec: Prob_Form}, the SBSs within a cluster serve users' requests using orthogonal RBs, therefore the user's will not experience any intra-cluster interference. Given the RBs are assigned to SBSs in a cluster, the matching decision of an SBS depends only on the matching policies of other clusters. As a result, the cost of an SBS is $\aggCost_{s}(\boldsymbol{\insCache}_s, \boldsymbol{\mathrm{\matchPolicy}}_s^{\cacheUpdateTime}, \boldsymbol{\mathrm{\matchPolicy}}_{\sbsClusterSet_{-s}}^{\cacheUpdateTime})$.
\vspace{-0.4cm}
\section{Profiling Popularity via Reinforcement Learning}
\label{sec: Pop_Prof}
The main objective of a caching strategy is to serve the user requests locally without accessing the cloud. An efficient caching strategy makes best use of the available cache by improving the instantaneous cache hits while minimizing the latency on the access and fronthaul link. Due to the varying spatio-temporal user demands, it is necessary to devise an adaptive algorithm to learn popularity profiles of classes in a decentralized manner. While classical caching mechanisms counts the frequency of content demands to obtain the actual content popularity, they ignore the impact of wireless channel conditions and scheduling. By observing the frequency of content demand, channel conditions and user scheduling, each SBS employs reinforcement learning to cache content. Such a latency-aware learning mechanism results in estimates close to the actual time average class popularity as time grows large which maximizes the average network utility. \\ 
\indent Let $\boldsymbol{\mathrm{\matchPolicy}}_{\sbsClusterSet_i}$ be the matrix of RBs' matching policies of cluster $i$ over time i.e., $\boldsymbol{\mathrm{\matchPolicy}}_{\sbsClusterSet} = [\boldsymbol{\mathrm{\matchPolicy}}_{\sbsClusterSet_1},\boldsymbol{\mathrm{\matchPolicy}}_{\sbsClusterSet_2} ..., ]$ where $\boldsymbol{\mathrm{\matchPolicy}}_{\sbsClusterSet_1}$ represents the matching matrix of the first cluster. According to \eqref{eq:sbss_utility}, the utility of an SBS depends on the caching policy and matching policy where the utility of an SBS is $\utility_{s}(\boldsymbol{\insCache}_s | \boldsymbol{\mathrm{\matchPolicy}}_{\sbsClusterSet}) = - \aggCost_{s}(\insCache_s, \boldsymbol{\mathrm{\matchPolicy}}_s)$. Based on the instantaneous matching policy, each SBS builds the probability distribution function of caching strategies to cache and serve users' requests. Each SBS caches content based on its own popularity profile and is independent of the contents cached by other SBSs in the same cluster. Although the matching policy varies over time, the utility function of an SBS depends on the matching policy of other SBSs due to inter-cluster interference and instantaneous channel conditions. The inter-cluster interference restraints the SBSs to calculate the exact utility function. In this regard, SBSs adopt regret learning technique to observe the utility based on the feedback from the UEs at each time $t$ that allow SBSs to explore all possible actions and learn optimal strategies \cite{Ref1:Reinforcement_Learning}. Here, each SBS maintains a measurement called regret per each action in which a positive regret indicates that the corresponding action should have been exploited in previous turns for higher payoff. To learn the class popularity, let $\actionSet_s = \{  a_{s,j} \}_{j \in \{1,\ldots, |\actionSet_s|\}}$ be the action space of SBS $s$ where $|\actionSet_s| =$ $\noOfContentClasses_s+\numCacheUp-1 \atopwithdelims ( ) \noOfContentClasses_s-1$ for $\numCacheUp$ cache updates and $\noOfContentClasses_s$ represents the number of popularity classes at SBS $s$. Here, an action $\chosenAction_s(t) \in \actionSet_s$ represents the choice of class(es) to be cached by SBS $s$ at time $t$. Let $\boldsymbol{\learningPop}_s(t) = [\learningPop_{s,j}(t)]_{j \in \{1,\ldots,A_s\} }$ be the probability distribution of SBS $s$ to select actions with $\learningPop_{s, j}(t) = \mathbb{P}(\chosenAction_s(t) = \chosenAction_{s,j})$, i.e., $\boldsymbol{\learningPop}_s(t)$ represents the mixed strategy of SBS $s$ at time $t$. When SBS employs an action $\chosenAction_{s}(t)$, it observes the utility based on the feedback from the UEs. Using the observed utility, the SBS minimizes the regret associated with action $\chosenAction_{s}(t)$. Let $\hat{\utility}_{s}(t)$ be the feedback of the sum of utilities from all associated users at SBS $s$, $\tilde{\utility}_{s}(t) = (\tilde{\utility}_{s,1}(t), ..., \tilde{\utility}_{s, |\actionSet_s|}(t))$ denote the estimated utility for each action at SBS $s$ at time $t$ and $\tilde{\boldsymbol{\regret}}_s(t) = (\tilde{\regret}_{s,1} (t), ..., \tilde{\regret}_{s,|\actionSet_s|} (t))$ denotes the estimated regret for each action. Based on the estimated regret, each SBS updates the caching strategy using the Boltzman's  Gibbs distribution as follows \cite{Ref18_CollaborativeCacheEdge3}:
\begin{equation}
\label{eq:Boltzamn}
\textstyle \gibbDist_{s,\chosenAction_{s,j}}(\tilde{\boldsymbol{\regret}}_s(t)) = \textstyle \frac {\textstyle \exp(\frac {\textstyle 1} {\textstyle \CacheTempCoeff_s} \textstyle \tilde{\regret}_{s,\chosenAction_{s,j}}^+(t))} {\textstyle \sum_{\forall \chosenAction_{s,j'} \in \actionSet_s} \textstyle \exp(\frac {\textstyle 1} {\textstyle \CacheTempCoeff_s} \textstyle \tilde{\regret}_{s,\chosenAction_{s,j'}}^+(t))}, \ \forall \chosenAction_{s,j} \in \actionSet_s,
\end{equation}
\noindent where $\CacheTempCoeff_s > 0$ is a temperature coefficient such that a small value of $\xi_s$ exploits actions yielding higher regrets while a higher value of $\CacheTempCoeff_s$ results in uniform distribution over the action set i.e., explores actions even that provided lower regrets \cite{Ref1:Clustering1}. The Boltzmann Gibbs distribution exploits the caching strategies with higher regrets while exploring caching strategies with lower regrets. The estimation of utility, regret and probability distribution over the action space at time $t$ is:
\vspace{-0.2cm}
\begin{align}
\label{eq:Regret}
\textstyle \tilde{\utility}_{s,\chosenAction_{s,j}}(t) &= \textstyle (1 - \learParam_1^s(t))\mathbb{1}_{\{\chosenAction_s = \chosenAction_{s,j}\}} \tilde{u}_{s,\chosenAction_{s,j}}(t-1) + \textstyle \learParam_1^s(t)\mathbb{1}_{\{\chosenAction_s = \chosenAction_{s,j}\}} \hat{\utility}_{s}(t), \nonumber \\
\textstyle \tilde{\regret}_{s,\chosenAction_{s,j}}(t) &= \textstyle (1 - \learParam_2^s(t)) \tilde{\regret}_{s,\chosenAction_{s,j}}(t-1) + \textstyle \learParam_2^s(t) (\tilde{\utility}_{s,\chosenAction_{s,j}}(t) - \textstyle \hat{\utility}_{s}(t)), \\
\textstyle \learningPop_{s,\chosenAction_{s,j}}(t) &= \textstyle (1 - \learParam_3^s(t)) \learningPop_{s,\chosenAction_{s,j}}(t-1) + \textstyle \learParam_3^s(t) \gibbDist_{s,\chosenAction_{s,j}}(\tilde{\boldsymbol{\regret}}_s(t)), \nonumber
\end{align}
\vspace{-1.0cm}

\noindent where the learning rates $\learParam_i^s(t)$ satisfies $\mathrm{lim}_{t \to \infty} \sum_{\tau = 1}^{t} \learParam_i^s(\tau) = \infty$, $\mathrm{lim}_{t \to \infty} \sum_{\tau = 1}^{t} \learParam_i^s(\tau) = \infty$ for $i \in \{1,2,3\}$ and $\mathrm{lim}_{t \to \infty} \sum_{\tau = 1}^{t} \frac {\learParam_{i+1}^s(t)} {\learParam_i^s(t)} = 0, \ i \in \{1, 2\}$ \cite{Ref1:Reinforcement_Learning}. \\
\indent Further, the cloud has the knowledge on the demands over the whole network. Based on this global knowledge, cloud learns the caching strategy $\boldsymbol{\learningPop}_c$ using steps similar to \eqref{eq:Regret} by modifying the action space i.e., $\actionSet_c = \contCatalogSet$ with $|\actionSet_c| = \contCatalog$. For each action, the corresponding utility and regret estimation at the cloud is $\utility_c(\actionSet_c) = \utility_{c}(\boldsymbol{\insCache}^{\contentClassesSet}, \boldsymbol{\frthaulPerUe})$ and $\tilde{\regret}_{c}(\actionSet_c) = \tilde{\regret}_{c,\boldsymbol{\insCache}^{\contentClassesSet}}$ respectively where $\boldsymbol{\insCache}^{\contentClassesSet} = [\boldsymbol{\insCache}^{\contentClassesSet_s}]_{s \in \sbsSize}$ represents the caches of all SBSs. \\
\indent To devise a combined popularity profile from the caching strategy of SBSs $\boldsymbol{\learningPop}_s$ and  cloud $\boldsymbol{\learningPop}_c$, the class popularity (the contents in same class having same popularity as defined in Sec. \ref{sssec:Caching_Strategy}) is used as follows:
\begin{align}
    \textstyle \popularity_{s,k_s} = \frac {1} {\sum_{k=1}^{\noOfContentClasses_s} \sum_{j=1}^{|\actionSet_s|} {\mathbb{1}_{\{\chi(a_{s,j}) = k\}}} \learningPop_{s,\chosenAction_{s,j}}(t) } \textstyle \sum_{j=1}^{|\actionSet_s|} \mathbb{1}_{\{\chi(a_{s,j}) = k_s\}} \learningPop_{s,\chosenAction_{s,j}}(t). \ \Longrightarrow \ \popularity_{s,f} = \frac {\popularity_{s,k_s}} {|k_s|}, f \in k_s, \label{eq:classProbDisSbs}
\end{align}
where $\mathbb{1}_{\{\chi(a_{s,j}) = k\}}$ indicates whether the class $k$ is cached or not at the SBS $s$ with the $j$-th action. Similarly the probability distribution of $f \in \contCatalogSet$ at the cloud is the same as cloud caching strategy i.e., $\popularity_{c,f} = \learningPop_{c,f}$. Based on the popularity profiles vector $\boldsymbol{\popularity}_{s}$ and $ \boldsymbol{\popularity}_{c}$, the combined popularity profile $\boldsymbol{\popularity}^{'}$ of $f$ and $k_s$ at SBS $s$ is given by:
\begin{equation}
\textstyle \popularity^{'}_{s, f} = (1-\locGlobTradeof) \frac {\popularity_{s,f}} { \popularity_{s,f} + \popularity_{c,f}} + \textstyle \locGlobTradeof \frac {\popularity_{c,f}} {\popularity_{s,f} + \popularity_{c,f}}, \ \Longrightarrow \ \popularity^{'}_{s, k_s} = \frac {1} {|k_s|} \sum_{f \in k_s} \popularity_{s, f},
\end{equation}
where $\locGlobTradeof$ is the local/global tradeoff control parameter such that $\locGlobTradeof = 1$ utilizes cloud driven popularity profile and $\locGlobTradeof = 0$ employs SBS driven popularity profile. Based on the popularity profile $\boldsymbol{\popularity}^{'}$, the combined caching strategy $\boldsymbol{\learningPop}'_s = [\learningPop'_s]_{a_s \in \actionSet_s}$ is given by:
\begin{equation}
\textstyle \boldsymbol{\learningPop}'_{s,a_{s,j}} = \frac {1} {\sum_{j=1}^{|\actionSet_s|} \sum_{k=1}^{\noOfContentClasses_s} 
 {\mathbb{1}_{\{\chi(a_{s,j}) = k\}}} \learningPop_{s,\chosenAction_{s,j}}(t)} \sum_{k=1}^{\noOfContentClasses_s} \mathbb{1}_{\{\chi(a_{s,j}) = k_s\}} \popularity^{'}_{s, k_s}.
\end{equation}
\vspace{-1.2cm}
\subsection{Convergence of RL based caching strategy}
In order to prove the convergence, consider the cost $\textstyle \boldsymbol{\cost_{su}}$ in \eqref{eq:Reward_eq1}. For a given scheduling policy $\boldsymbol{\schedVar_{su}}(t)$, the cost is given as:
\begin{equation*}
    \textstyle \boldsymbol{\cost_{su}(\boldsymbol{x}(t))} = - \textstyle \frac{\textstyle \boldsymbol{\transBits_{su}}(t)} {\textstyle \boldsymbol{\schedVar_{su}}(t) \log_2 (1 + \textstyle \boldsymbol{x}_{su}(t)) + c },
\end{equation*}
\vspace{-0.1cm}
where $x_{su}$ is the SINR of user $u$. \\
\textit{Proof:} See Appendix A. 
\vspace{-0.4cm}
\begin{theorem}
As $t \to \infty$, Alg. \ref{algo:algo1} converges to the stationary distribution $\Pi_s = (\Pi_{s,a_s}, \forall a_s \in \actionSet_s)$ with, 
\begin{equation*}
    \Pi_{s,a_s} = \textstyle \frac {\textstyle \exp(\frac{\textstyle 1} {\textstyle \CacheTempCoeff_s} \textstyle \tilde{\utility}_{s,\chosenAction_{s}})} {\textstyle \sum_{\chosenAction_{s}^{'} \in \actionSet_s} \textstyle \exp(\frac {\textstyle 1} {\textstyle \CacheTempCoeff_s} \textstyle \tilde{\utility}_{s,\chosenAction_{s}^{'}})}.
\end{equation*}
As a result, as $\CacheTempCoeff_s \to 0$,
\begin{equation}
\textstyle \lim_{\CacheTempCoeff_s \to 0} \Pi_{s,a_s}^{(\CacheTempCoeff_s)} = \textstyle \Pi_{s,a_s}^{(0)} = \textstyle \begin{cases} \textstyle \frac {\textstyle 1} {\textstyle |\actionSet_s^{*}|} \ \ \ \ & \mathrm{\text{if}} \textstyle \ a_s \in \actionSet_s^{*}, \\
\textstyle 0 \ \ \ \ & \mathrm{\text{if}} \textstyle \ a_s \notin \actionSet_s^{*}, 
\end{cases} 
\end{equation}
where $\actionSet_s^{*}$ is the global optimal solution.
\end{theorem}
\vspace{-0.2cm}
\begin{IEEEproof}
Let $a_s \in \actionSet_s^{*}$. Therefore $\tilde{\utility}_{s,\chosenAction_{s}} > \tilde{\utility}_{s,\chosenAction_{s}^{'}}, \forall \chosenAction_{s}^{'} \notin \actionSet_s^{*}$. Then,
\begin{align*}
    \textstyle \Pi_{s,a_s}^{(0)} = & \lim_{\CacheTempCoeff_s \to 0} \frac { \exp(\frac{ 1} { \CacheTempCoeff_s} \tilde{\utility}_{s,\chosenAction_{s}})} { \sum_{\chosenAction_{s}^{'} \in \actionSet_s} \exp(\frac { 1} { \CacheTempCoeff_s} \tilde{\utility}_{s,\chosenAction_{s}^{'}})} = \lim_{\CacheTempCoeff_s \to 0} \frac { 1} { |\actionSet_s^{*}| + \sum_{\chosenAction_{s}^{'} \notin \actionSet_s^{*}} \exp(\frac{ 1} { \CacheTempCoeff_s} (\tilde{\utility}_{s,\chosenAction_{s}^{'}} - \tilde{\utility}_{s,\chosenAction_{s}}))} = \frac { 1} { |\actionSet_s^{*}|}.
\end{align*}
The above theorem states that $\CacheTempCoeff_s > 0$ does not ensure optimal solution under stationary popularity distribution. However as $\CacheTempCoeff_s \to 0$, the algorithm converges to the optimal solution.
\end{IEEEproof}
\vspace{-0.4cm}
\section{Optimal User Scheduling and RB Matching via Lyapunov Framework}
Due to the time-varying user demands, channel conditions and resources, \eqref{eq: Obj_Func} becomes a stochastic optimization problem. Lyapunov optimization framework is a powerful tool for solving stochastic optimization problem \eqref{eq: Obj_Func} while ensuring queue stability \eqref{eq:Obj_Func10}. Stochastic optimization approach splits the original problem into multiple subproblems by utilizing the DPP approach of the lyapunov framework allowing individual SBSs to make their own decisions with low coordination overhead. Inspired from lyapunov DPP framework, this work employs lyapunov framework by decoupling \eqref{eq: Obj_Func} into user scheduling, resource allocation and fronthaul allocation subproblems and solving them individually. In this respect, SBSs optimizes user scheduling and resource allocation  while cloud optimizes the fronthaul allocation. Since the SBSs cannot transmit and receive simultaneously, \eqref{eq: Obj_Func} is optimized separately at the cloud and SBSs. However, due to the probabilistic non-linear constraint \eqref{eq:Obj_Func3} and \eqref{eq:Obj_Func4}, \eqref{eq: Obj_Func} is not tractable and challenging to solve. To derive a tractable solution, the probabilistic constraints in \eqref{eq:Obj_Func3}, \eqref{eq:Obj_Func4} are transformed into deterministic constraints using Markov's inequality \cite{Ref2:Markov_Ref} such that $\mathbbm{P} \big\{ \frac {\dataQueue_{su}(t)} {\bar{\packetArrRate}_{su}^f} > \qosReq_{sq_{uf}} \big\} \leq \frac {\mathbbm{E}[\dataQueue_{su}(t)]} {\bar{\packetArrRate}_{su}^f {\qosReq_{sq_{uf}}}}$ and $\mathbbm{P} \big\{ \packetSize_f \dataQueue_{cs}^f(t) > \qosReq_{sq_{uf}} \big\} \leq \frac {\mathbbm{E}[\dataQueue_{cs}(t)] \packetSize_f} {\qosReq_{sq_{uf}}}$ where $\mathbbm{E}[.]$ represents the expected value. 
In addition, due to the non-convex nature of the objective \eqref{eq:Obj_Func1} 
, new variables $[E_s]_{\forall s \in \sbsSet}$ and $B$ are introduced for \eqref{eq:sbss_utility} and \eqref{eq:cloud_utility} respectively which represent the maximum latency experienced by an SBS and the fronthaul in serving a user request. Let $\bar{E}_s = \frac {1} {\cacheUpdateTime} \sum_{t=0}^{\cacheUpdateTime - 1} E_s(t)$ and $\bar{B} = \frac {1} {\cacheUpdateTime} \sum_{t=0}^{\cacheUpdateTime - 1} B(t)$ be the time-average value of $E_s(t)$ and $B(t)$ respectively. Using the Markov inequality on \eqref{eq:Obj_Func3} and  \eqref{eq:Obj_Func4} assuming the fixed optimal caching policy in Sec. \ref{sec: Pop_Prof} and fixed clustering, the transformed cost optimization problem \eqref{eq: Obj_Func} is:
\vspace{-1.0cm}

\begin{subequations}
\label{eq: Lyapunov_Opt2}
\begin{align}
    \underset{\textstyle \boldsymbol{\mathrm{\matchPolicy}}(t), \boldsymbol{\schedVar}(t), \boldsymbol{\frthaulPerUe}(t), \boldsymbol{E}(t), B(t), \ \forall t}{\text{minimize}} \ \ \ & \textstyle \frac {1} {| \sbsClusterSet |} \sum_{\forall \sbsClusterSet_i \in \sbsClusterSet} \textstyle \bigg( \frac {1} {| \sbsSet_{\sbsClusterSet_i} |} \sum_{\forall s \in \sbsClusterSet_i} \textstyle \bar{E}_s \bigg) + \bar{B}, \label{eq:Lyapunov_Opt21} \\
    \text{subject to} \ \ & \textstyle E_{s}(t) \geq \cost_{su}^{f}(\boldsymbol{\insCache}_s^{*}(t), \boldsymbol{\mathrm{\matchPolicy}}(t)) \ \ \forall s \in \sbsSet, t, \label{eq:Lyapunov_Opt22} \\
    \ \ & \textstyle B(t) \geq \cost_{cs}^{f}( \boldsymbol{\insCache}_s^{*}(t), \boldsymbol{\frthaulPerUe}(t)), \ \ \forall t, \label{eq:Lyapunov_Opt23} \\
    \ \ & \textstyle \bar{\dataQueue}_{su}^f \leq \epsilon_u \bar{\packetArrRate}_{su}^f {\qosReq_{sq_{uf}}}, \ \ \forall s \in \sbsSet, u \in \ueSet, f \in \contCatalogSet, \label{eq:Lyapunov_Opt2} \\
    \ \ & \textstyle \bar{\dataQueue}_{cs}^f \leq \frac {\epsilon_s {\qosReq_{sq_{uf}}}} {\packetSize_f}, \ \ \forall s \in \sbsSet, f \in \contCatalogSet, \label{eq:Lyapunov_Opt3} \\
    \ \ & \textstyle \eqref{eq:Obj_Func3} - \eqref{eq:Obj_Func10} \label{eq:Lyapunov_Opt6}.
\end{align}
\end{subequations}
\indent To satisfy \eqref{eq:Lyapunov_Opt22} - \eqref{eq:Lyapunov_Opt23}, we associate a set of deficit queues $\virtQueue_{su}^f(t)$, and $\virtQueue_{cs}^f(t)$, which tracks the corresponding user queues $\dataQueue_{su}^f(t)$ and $\dataQueue_{cs}^f(t)$ and determines the current lag of latency experienced by SBS and fronthaul behind the target value. The evolution of these queues are:
\begin{align}
\textstyle \virtQueue_{su}^f(t+1) & = \max[\virtQueue_{su}^f(t) + \dataQueue_{su}^f(t+1) - \epsilon_u \bar{\packetArrRate}_{su}^f {\qosReq_{sq_{uf}}}, 0], \ \label{eq:Deficit_queue1} \\
\textstyle \virtQueue_{cs}^f(t+1) & = \max[\virtQueue_{cs}^f(t) + \dataQueue_{cs}^f(t+1) - \frac {\epsilon_u {\qosReq_{sq_{uf}}}} {\packetSize_f} , 0]. \ \label{eq:Deficit_queue2}
\vspace{-0.3cm}
\end{align}
For notational simplicity, let 
$\boldsymbol{\mathrm{\ueRbsDefVec}}(t) \overset{\Delta}{=} \big([\boldsymbol{\mathrm{\dataQueue}}_s(t)]_{s \in \sbsSet}, \boldsymbol{\mathrm{\dataQueue}}_c(t), \boldsymbol{\rbQueueEvo}(t), [\boldsymbol{\mathrm{\virtQueue}}_s(t)]_{s \in \sbsSet}, \boldsymbol{\mathrm{\virtQueue}}_c(t) \big)$ be the combined queue vector. Let the lyapunov function be defined by $\lyapunov(\boldsymbol{\mathrm{\ueRbsDefVec}}(t)) = \frac {1} {2} \lVert \boldsymbol{\mathrm{\ueRbsDefVec}}(t) \rVert^2$, then the one-slot lyapunov drift $\Delta \lyapunov = \lyapunov(\boldsymbol{\mathrm{\ueRbsDefVec}}(t+1)) - \lyapunov(\boldsymbol{\mathrm{\ueRbsDefVec}}(t))$ is \cite{Ref26_Intro_Stochastic}:
\vspace{-0.1cm}
\begin{align}
\textstyle \Delta \lyapunov = \frac {\textstyle 1} {\textstyle 2} \bigg[ & \sum_{s \in \sbsSet} \sum_{u \in \ueSet}\bigg\{ \big( \boldsymbol{\dataQueue}_{su}(t+1) - \boldsymbol{\dataQueue}_{su}(t) \big)\IEEEauthorrefmark{2}\big( \boldsymbol{\dataQueue}_{su}(t+1) - \boldsymbol{\dataQueue}_{su}(t) \big) \bigg\} + \sum_{s \in \sbsSet} \bigg\{ \rbQueueEvo_s^2(t+1) - \nonumber \\ 
 \textstyle \rbQueueEvo_s^2(t) & \bigg\} + \sum_{s \in \sbsSet} \bigg\{ \big( \boldsymbol{\dataQueue}_{cs}(t+1) - \boldsymbol{\dataQueue}_{cs}(t) \big)\IEEEauthorrefmark{2}\big( \boldsymbol{\dataQueue}_{cs}(t+1) - \boldsymbol{\dataQueue}_{cs}(t) \big) \bigg\} + \sum_{s \in \sbsSet} \sum_{u \in \ueSet}\bigg\{ \big( \boldsymbol{\virtQueue}_{su}(t+1) - \nonumber \\
\textstyle \boldsymbol{\virtQueue}_{su}(t) & \big)\IEEEauthorrefmark{2} \big( \boldsymbol{\virtQueue}_{su}(t+1) - \boldsymbol{\virtQueue}_{su}(t)\big) \bigg\} + \sum_{s \in \sbsSet} \bigg\{ \big( \boldsymbol{\virtQueue}_{cs}(t+1) - \boldsymbol{\virtQueue}_{cs}(t) \big)\IEEEauthorrefmark{2}\big( \boldsymbol{\virtQueue}_{cs}(t+1) - \boldsymbol{\virtQueue}_{cs}(t)\big) \bigg\} \bigg].
\end{align}
\theoremstyle{definition}
\begin{lemma}{}
Under a feasible policy, we have the following inequality:
\vspace{-0.3cm}
\begin{align*}
\textstyle \Delta(L) \leq & \ C + \sum_{s \in \sbsSet} \sum_{u \in \ueSet} \sum_{f \in \contCatalogSet} \bigg[ \dataQueue_{su}^f(t) \Big( \packetArrRate_{su}^f(t) - \sum_{m=1}^{\rbsSize} \rate_{su}^{(m,f)}(t) \Big) + \virtQueue_{su}^f(t) \Big( \dataQueue_{su}^f(t) - \sum_{m=1}^{\rbsSize} \rate_{su}^{(m,f)}(t) + \nonumber \\
\textstyle & \packetArrRate_{su}^f(t) - \epsilon_u \bar{\packetArrRate}_{su}^f \qosReq_{s\ueRequests_uf} \Big) \bigg] + \sum_{s \in \sbsSet} \sum_{f \in \contCatalogSet} \bigg[ \dataQueue_{cs}^f(t) \Big( \frac {\textstyle 1} {\textstyle \packetSize_f} \sum_{u \in \ueCoverage_s(t)} \mathbbm{1}_{\{q_u(t) \not\in \insCache_s(t) \} } - (1-\sbsTransDur) \frthaulPerUe_{s}^f \frthaulCap \Big) + \nonumber
\end{align*}
\begin{align*}
& \textstyle \virtQueue_{cs}^f(t) \Big( \dataQueue_{cs}^f(t) - (1-\sbsTransDur) \frthaulPerUe_{s}^f \frthaulCap + \frac {1} {\packetSize_f} \sum_{u \in \ueCoverage_s(t)} \mathbbm{1}_{\{q_u(t) \not\in \insCache_s(t) \}} - \frac {\textstyle \epsilon_s \qosReq_{s\ueRequests_uf}} {\textstyle \packetSize_f} \Big) \bigg] + \nonumber \\
& \textstyle \sum_{s \in \sbsSize} \bigg[ \rbQueueEvo_{su}(t) \Big( \rbsArrRate_s(t) - \sum_{u \in \ueCoverage_s(t)} |\matchPolicy_{su}(t)| \Big) \bigg],
\end{align*}
where $C$ is the uniform bound on $\frac {1} {2} \Big[ \Big( \packetArrRate_{su}^f(t) - \sum_{m=1}^{\rbsSize} \rate_{su}^{(m,f)}(t) \Big)^2 + \Big( \rbsArrRate_s(t) - \sum_{u \in \ueCoverage_s(t)} |\matchPolicy_{su}(t)| \Big)^2 + \Big( \frac {1} {\packetSize_f} \sum_{u \in \ueCoverage_s(t)} \mathbbm{1}_{\{q_u(t) \not\in \insCache_s(t) \} } - (1-\sbsTransDur) \frthaulPerUe_{s}^f \frthaulCap \Big)^2 + \Big( \epsilon_u \bar{\packetArrRate}_{su}^f \qosReq_{s\ueRequests_uf} - \dataQueue_{su}^f(t) + \sum_{m=1}^{\rbsSize} \rate_{su}^{(m,f)}(t) - \packetArrRate_{su}^f(t) \Big)^2 + \Big( \frac {\epsilon_s \qosReq_{s\ueRequests_uf}} {\packetSize_f} - \dataQueue_{cs}^f(t) + (1-\sbsTransDur) \frthaulPerUe_{s}^f \frthaulCap - \frac {1} {\packetSize_f} \sum_{u \in \ueCoverage_s(t)} \mathbbm{1}_{\{q_u(t) \not\in \insCache_s(t) \} } \Big)^2  \Big]$.
\end{lemma}
\textit{Proof:} See Appendix B. 

The one-slot conditional Lyapunov DPP is defined as $\Delta(\lyapunov) \overset{\Delta_s}{=} \mathbb{E}\{\lyapunov(\boldsymbol{\mathrm{\ueRbsDefVec}}(t+1)) - \lyapunov(\boldsymbol{\mathrm{\ueRbsDefVec}}(t)) | \boldsymbol{\mathrm{\ueRbsDefVec}}(t)\} + \utilDelayTradeoff \bigg[ \frac {1} {| \sbsClusterSet |} \sum_{\forall \sbsClusterSet_i \in \sbsClusterSet} \Big( \frac {1} {| \sbsSet_{\sbsClusterSet_i} |} \sum_{\forall s \in \sbsClusterSet_i} E_s(t) \Big) + B(t) \bigg]$ where $\utilDelayTradeoff \geq 1$ is a tunable parameter for controlling the tradeoff between optimality of the solution and the queue length and $\Delta L \overset{\Delta}{=} \mathbb{E}\{\lyapunov(\boldsymbol{\mathrm{\ueRbsDefVec}}(t+1)) - \lyapunov(\boldsymbol{\mathrm{\ueRbsDefVec}}(t)) | \boldsymbol{\mathrm{\ueRbsDefVec}}(t)\}$ is the conditional lyapunov drift. The optimization variables $\boldsymbol{\mathrm{\matchPolicy}}(t)$, $\boldsymbol{\mathrm{\schedVar}}(t)$ and $\boldsymbol{E}(t)$ for SBSs are captured in a variable $\varphi(t)$ such that $\boldsymbol{\varphi}(t) = [\boldsymbol{\mathrm{\matchPolicy}}(t), \boldsymbol{\schedVar}(t), \boldsymbol{E}(t)]$. Similarly, the optimization variables for the cloud $\vartheta(t) = [\boldsymbol{\frthaulPerUe}(t), B(t)]$. Let $\boldsymbol{\mathrm{\matchPolicy}}_s^{\mathrm{avg}}(t) = \frac {1} {t} \sum_{\tau=0}^{t-1} \boldsymbol{\mathrm{\matchPolicy}}_s(\tau)$, $\boldsymbol{\frthaulPerUe}_s^{\mathrm{avg}}(t) = \frac {1} {t} \sum_{\tau=0}^{t-1} \boldsymbol{\frthaulPerUe}_s(\tau)$, $\boldsymbol{\schedVar}_s^{\mathrm{avg}}(t) = \frac {1} {t} \sum_{\tau=0}^{t-1} \boldsymbol{\schedVar}_s(\tau)$, $\boldsymbol{E}_s^{\mathrm{avg}}(t) = \frac {1} {t} \sum_{\tau=0}^{t-1} \boldsymbol{E}_s(\tau)$ and $\boldsymbol{B}^{\mathrm{avg}}(t) = \frac {1} {t} \sum_{\tau=0}^{t-1} \boldsymbol{B}(\tau)$ respectively be the current running time average of the matching policy at SBS $s$, percentage of fronthaul link at the cloud, scheduling matrix and convexification variables for the objective function. Further, let $V \Big[ \nabla_{\boldsymbol{\varphi}_s} \big\{ \frac {1} {| \sbsClusterSet |} \sum_{\forall \sbsClusterSet_i \in \sbsClusterSet} \big( \frac {1} {| \sbsSet_{\sbsClusterSet_i} |} \sum_{\forall s \in \sbsClusterSet_i} E_s(t) \big) \big\} + \nabla_{\boldsymbol{\vartheta}} B(t) \Big]$ be the penalty introduced based on the combined queues, then the problem aims to solve the upper bound of the drift DPP i.e., 
\vspace{-0.3cm}
\begin{align*}
\textstyle C & \ + \sum_{s \in \sbsSet} \sum_{u \in \ueSet} \sum_{f \in \contCatalogSet} \bigg[ \mathbbm{E} \Big\{ \dataQueue_{su}^f(t) \Big( \packetArrRate_{su}^f(t) - \sum_{m=1}^{\rbsSize} \rate_{su}^{(m,f)}(t) \Big) | v(t) \Big\} + \mathbbm{E} \Big\{ \virtQueue_{su}^f(t) \Big( \dataQueue_{su}^f(t) - \sum_{m=1}^{\rbsSize} \rate_{su}^{(m,f)}(t) \nonumber \\
& \textstyle + \packetArrRate_{su}^f(t) - \epsilon_u \bar{\packetArrRate}_{su} \qosReq_{s\ueRequests_uf} \Big) | v(t) \Big\} \bigg] + \sum_{s \in \sbsSet} \Big[ \mathbbm{E} \Big\{ \rbQueueEvo_{s}(t) \Big( \rbsArrRate_s(t) - \sum_{u \in \ueCoverage_s(t)} |\boldsymbol{\matchPolicy}_{su}(t)| \Big) | v(t) \Big\} \Big] + \nonumber \\
& \textstyle \sum_{s \in \sbsSet} \sum_{f \in \contCatalogSet} \bigg[ \mathbbm{E} \Big\{ \dataQueue_{cs}^f(t) \Big( \frac {\textstyle 1} {\textstyle \packetSize_f} \sum_{u \in \ueCoverage_s(t)} \mathbbm{1}_{\{q_u(t) \not\in \insCache_s(t) \} } - (1-\sbsTransDur) \frthaulPerUe_{s}^f \frthaulCap \Big) | v(t) \Big\} + \mathbbm{E} \Big\{ \virtQueue_{cs}^f(t) \Big( \dataQueue_{cs}^f(t) \nonumber \\
& \textstyle - (1-\sbsTransDur) \frthaulPerUe_{s}^f \frthaulCap + \frac {1} {\packetSize_f} \sum_{u \in \ueCoverage_s(t)} \mathbbm{1}_{\{q_u(t) \not\in \insCache_s(t) \}} - \frac {\textstyle \epsilon_s \qosReq_{s\ueRequests_uf}} {\textstyle \packetSize_f} \Big) + \nonumber \\
& \textstyle V \Big[ \nabla_{\boldsymbol{\varphi}_s} \big\{ \frac {1} {| \sbsClusterSet |} \sum_{\forall \sbsClusterSet_i \in \sbsClusterSet} \big( \frac {\textstyle 1} {\textstyle | \sbsSet_{\sbsClusterSet_i} |} \sum_{\forall s \in \sbsClusterSet_i} E_s(t) \mathbbm{E}[\varphi_s(t)|\ueRbsDefVec(t)] \big) \big\} + \nabla_{\boldsymbol{\vartheta}} B(t) \mathbbm{E}[\vartheta(t)|\ueRbsDefVec(t)] \Big]. \nonumber
\end{align*}

\noindent The terms $C$, $\dataQueue_{su}^f(t) \packetArrRate_{su}^f(t)$, $\rbQueueEvo_{su}(t) \rbsArrRate_s(t)$, $\virtQueue_{su}^f(t) \dataQueue_{su}^f(t)$, $\virtQueue_{su}^f(t) \packetArrRate_{su}^f(t)$, $\virtQueue_{cs}^f(t) \dataQueue_{cs}^f(t)$, $\virtQueue_{cs}^f(t) \\ \frac {1} {\packetSize_f} \sum_{u \in \ueCoverage_s(t)} \mathbbm{1}_{\{q_u(t) \not\in \insCache_s(t) \}}$ and $\dataQueue_{cs}^f(t) \frac {1} {\packetSize_f} \sum_{u \in \ueCoverage_s(t)} \mathbbm{1}_{\{q_u(t) \not\in \insCache_s(t) \} }$ are independent from the decision variables $\mathrm{\boldsymbol{\matchPolicy}}_s(t)$ and $\boldsymbol{\frthaulPerUe}$ respectively, and therefore, can be neglected. As a result, the global objective is to minimize the following term:
\vspace{-0.9cm}

\begin{align}
\label{eq:opt_eq} 
 & \ \textstyle \overbrace{ V \Big[ \bigg\{ \frac {\textstyle 1} {\textstyle | \sbsClusterSet |} \sum_{\forall \sbsClusterSet_i \in \sbsClusterSet} \bigg( \frac {\textstyle 1} {\textstyle | \sbsSet_{\sbsClusterSet_i} |} \sum_{\forall s \in \sbsClusterSet_i} \nabla_{\boldsymbol{\varphi}_s} E_s(t) \varphi_s(t) \bigg) \bigg\} + \nabla_{\boldsymbol{\vartheta}} B(t) \vartheta(t) \Big]}^{\text{penalty}} - \sum_{\forall s \in \sbsSet}  \overbrace{\bigg( \rbQueueEvo_{s}(t) \sum_{u \in \ueCoverage_s(t)} |\matchPolicy_{su}(t)|\bigg)}^{\text{Impact of resource queue and matching}} \nonumber \\
  & \ \textstyle - \sum_{u \in \ueSet} \sum_{f \in \contCatalogSet} \overbrace{\bigg( \dataQueue_{su}^f(t) \sum_{m=1}^{\rbsSize} \rate_{su}^{(m,f)}(t)}^{\text{QSI and CSI at SBSs}} + \overbrace{\virtQueue_{su}^f(t) \Big( \sum_{m=1}^{\rbsSize} \rate_{su}^{(m,f)}(t) + \epsilon_u \bar{\packetArrRate}_{su}^f \qosReq_{s\ueRequests_uf} \Big) 
\bigg)}^{\text{Impact of virtual queue, QSI, CSI and auxiliaries at SBSs}} \nonumber \\
& \ \textstyle - \sum_{s \in \sbsSet} \sum_{f \in \contCatalogSet} \overbrace{\bigg( \virtQueue_{cs}^f(t) \Big( \frthaulPerUe_{s}^f \frthaulCap (1-\sbsTransDur) + \frac {\textstyle \epsilon_s \qosReq_{s\ueRequests_uf}} {\textstyle \packetSize_f} \Big)
}^{\text{Impact of QSI and auxiliaries at the cloud}} + \overbrace{(1-\sbsTransDur) \dataQueue_{cs}^f(t) \frthaulPerUe_{s}^f \frthaulCap}^{\text{QSI and CSI at cloud}} \bigg).
\end{align}

\noindent Although the matching policy of SBS and auxiliary variables have been decoupled, the SBSs are unaware of the matching policy in other clusters due to the absence of knowledge of coupling. As a result, the achievable rate is not completely decoupled. To address this issue, we propose a sampling based learning algorithm to estimate the interference at SBSs. Let $\boldsymbol{\tilde{I}}_{su}(z(t)) = [\tilde{I}_{su}^m(z(t))]_{u \in \mathcal{U}}$ be the matrix of estimated interference matrix at SBS $s$ at time $t$, then the time-average estimation of interference is given as:
\begin{equation}
\label{eq:EstimatedInterference1}
\textstyle \boldsymbol{\tilde{I}}_{su}(t,z(t)) = \nu \boldsymbol{\tilde{I}}_{su}(t,z(t-1)) + (1 - \nu) ( \boldsymbol{I}_{su}(t, z(t-1) - \boldsymbol{\tilde{I}}_{su}(t,z(t-1))).
\end{equation}
By replacing the interference by the estimated interference, all the terms in \eqref{eq:opt_eq} are decoupled and can be solved individually. Further, the instantaneous rate at the SBS $s$ with optimized cache $\insCache_s^{*}(t)$ in terms of estimated interference is:
\begin{equation}
\label{eq:New_Rate}
\textstyle \rate_{su}^{(m,f)}(t) = \sbsTransDur \bwPerRB \schedVar_{su}^f(t) \log_2 \Big( 1 + \frac { \textstyle \matchPolicy_{su}^m(\boldsymbol{\mathrm{Z}}(t), \boldsymbol{\insCache}_s^{*}(t)) \txPower_s^m(t) \channel_s^m(t)} {\textstyle \noise + \tilde{I}_{su}^m(\boldsymbol{\mathrm{Z}}(t))} \Big).
\end{equation}
\subsubsection{Determining matching policy}
The matching policy of SBS is determined by separating \eqref{eq:opt_eq} into sub problems which represents the impact of $\rbQueueEvo_{s}(t)$, $\rate_{su}^{(m,f)}(t)$, $\dataQueue_{su}^f(t)$ and $\virtQueue_{su}^f(t)$. Each sub problem provides the means to find resource allocation and user scheduling, as:
\begin{subequations}
\label{eq: sbsOptimization}
\begin{align}
    \underset{\textstyle \boldsymbol{\mathrm{\matchPolicy}}_{s}(t), \boldsymbol{\schedVar}(t), \boldsymbol{E}(t)}{\text{maximize}} \ \ \ & \textstyle \rbQueueEvo_{s}(t) \sum\limits_{u \in \ueCoverage_s(t)}^{m \in \mathcal{M}_s(t)} \matchPolicy_{su}^m(t) + \sum_{u \in \ueSet} \sum_{f \in \contCatalogSet} \bigg( \big\{ \dataQueue_{su}^f(t) + \virtQueue_{su}^f(t) \big\} \nonumber \\ 
    & \textstyle \sum_{m=1}^{\rbsSize} \rate_{su}^{(m,f)}(t) + \virtQueue_{su}^f(t) \epsilon_u \bar{\packetArrRate}_{su}^f \qosReq_{s\ueRequests_uf} \bigg) - V E_s(t), \ \label{eq:sbsOptimization1} \\
    \text{subject to} \ \ & \eqref{eq: queue_eq1}, \eqref{eq: queue_eq2}, \eqref{eq:Obj_Func4} - \eqref{eq:Obj_Func6}, \eqref{eq:Obj_Func8}, \eqref{eq:Lyapunov_Opt22}, \ \ \ \forall s, u, m. \label{eq:sbsOptimization2}
\end{align}
\end{subequations}

\noindent The optimization problem is non-convex due to: 1) achievable rate is a non-convex function of matching variable and scheduling variable as per \eqref{eq:New_Rate}, 2) \eqref{eq:Lyapunov_Opt22} is non-convex because the matching variable and the scheduling variable are integers. To address the integer nature of the matching and scheduling variables, we assume a linear relaxation such that $\boldsymbol{\matchPolicy}_{su}(t) \in [0,1]$ and $\schedVar_{su}^f(t) \in [0,1]$. To recast the achievable rate in \eqref{eq:sbsOptimization1} and \eqref{eq:Lyapunov_Opt22}, lets assume that the transmit power is equal over all RBs i.e., $\txPower_s^m(t) = \frac {p_s^{\mathrm{max}}} {\sum_{u \in \ueCoverage_s(t)} \sum_{m'=1}^{\mathcal{M}_s(t)} \matchPolicy_{su}^{m'}(t)}$. The achievable rate is:
\begin{equation}
\label{eq:Rate_pwr}
    \textstyle \rate_{su}^{(m,f)}(t) = \sbsTransDur \bwPerRB \schedVar_{su}^f(t) \log_2 \Big( 1 + \frac { \matchPolicy_{su}^m(\boldsymbol{\mathrm{Z}}(t), \boldsymbol{\insCache}_s^{*}(t)) \zeta_{su}^{m}(t) p_s^{\mathrm{max}} } { \sum_{u \in \ueCoverage_s(t)} \sum_{m'=1}^{\mathcal{M}_s(t)} \matchPolicy_{su}^{m'}(t)} \Big),  
\end{equation}
\indent where $\zeta_{su}^{m}(t) = \frac {\channel_{su}^m(t)} {\noise + \tilde{I}_{su}^m(\boldsymbol{\mathrm{Z}}(t))}$ is the Channel-to-Interference plus Noise (CINR) between SBS $s$ and user $u$ over RB $m$. The achievable rate in \eqref{eq:Rate_pwr} can be represented as:
\begin{align}
    \textstyle \rate_{su}^{(m,f)}(t) = & \sbsTransDur \bwPerRB \schedVar_{su}^f(t) \Big[ \log_2  \Big( \sum_{u \in \ueCoverage_s(t)} \boldsymbol{\rbRepresentation}^m(t) \matchPolicy_{su}^m(\boldsymbol{\mathrm{Z}}(t), \boldsymbol{\insCache}_s^{*}(t)) \Big) - \log_2 \Big( \sum_{u \in \ueCoverage_s(t)} \sum_{m'=1}^{\rbsSize_s(t)} \matchPolicy_{su}^{m'}(t) \Big) \Big], \label{eq:Rate_Cvx}
\end{align}
\indent where $\boldsymbol{\rbRepresentation}^m(t) = [1 ... 1_{m-1} \ 1+\zeta_{su}^{m}(t) p_s^{\mathrm{max}} \ 1 ... 1_{\rbsSet(t)} ]$. Note that the achievable rate is a product of scheduling variable and difference of concave functions with respect to matching variables.
\subsubsection{Approximation of achievable rate per RB}
To approximate the difference of concave functions in \eqref{eq:Rate_Cvx}, a new constraint is imposed as:
\vspace{-0.2cm}
\begin{align}
\textstyle & \underbrace{x_{su}^m(t) - \log_2 \Big( \sum_{u \in \ueCoverage_s(t)} \boldsymbol{\rbRepresentation}^m(t) \matchPolicy_{su}^m(\boldsymbol{\mathrm{Z}}(t),  \boldsymbol{\insCache}_s^{*}(t)) \Big)}_{f_0(b_{su}^m)} + \underbrace{\log_2 \Big( \sum_{u \in \ueCoverage_s(t)} \sum_{m'=1}^{\rbsSize_s(t)} \matchPolicy_{su}^{m'}(t) \Big)}_{g_0(b_{su}^m)} \leq 0. \label{eq:Rate1Constraint}
\end{align}
\indent For any variable $x \in \mathbb{R}$ and convex function $g(x)$, the convex form of $g(x)$ at iteration $k$ is:
\vspace{-0.2cm}
\begin{equation}
    \hat{g}(x;x_k) \overset{\Delta}{=} g(x_k) + \bigtriangledown g(x_k) (x - x_k),
\vspace{-0.2cm}
\end{equation}
\indent where $\bigtriangledown g(x_k)$ represents the gradient of  $g(x_k)$. Using an approximated convex method [45]. $g_0(b_{su}^m)$ in \eqref{eq:Rate1Constraint} is replaced with its first order approximation at iteration $k$ as follows:
\begin{align}
\textstyle g_0(b_{su}^m;b_{su}^m(k)) = g_0(b_{su}^m, b_{su}^m(k)) + \frac {1} {\sum_{u \in \ueCoverage_s(t)} \sum_{m'=1}^{\rbsSize_s(t)} \matchPolicy_{su}^{m'}(k)} \sum_{u \in \ueCoverage_s(t)} \sum_{m'=1}^{\rbsSize_s(t)} \bigg( \matchPolicy_{su}^{m'} - \matchPolicy_{su}^{m'}(k) \bigg). 
\end{align}
\noindent Using the analogy that epigraph of $f$ is convex iff the region $f(x)$ is convex set and $2 xy = (x+y)^2 - x^2 - y^2$, \eqref{eq:Rate_Cvx} can be rewritten as follows:
\begin{align}
    \textstyle \rate_{su}^{(m,f)}(t) = \frac {\sbsTransDur \bwPerRB} {2} \Bigg[ \underbrace{\bigg\{ \schedVar_{su}^f(t) + x_{su}^m(t) \bigg\}^2}_{f_1(Y_{su}^f, x_{su}^m)} - \underbrace{ \bigg\{ \schedVar_{su}^f(t)^2 + x_{su}^m(t)^2 \bigg\} }_{g_1(Y_{su}^f, x_{su}^m)} \Bigg].  \ \ \label{eq:rate_affine}
\end{align}
\noindent Note that the achievable rate has been written as a difference of two convex functions which will be solved using the difference of convex function technique i.e., concave-convex procedure (CCP) \cite{Ref1:CCP}. By replacing $g_0(Y_{su}^f, x_{su}^m)$ in \eqref{eq:rate_affine} with its first order approximation at iteration $k$ as:  
\begin{align}
    \textstyle g_0(Y_{su}^f,x_{su}^m;Y_{su}^f(k),x_{su}^m(k)) & = g_0(Y_{su}^f(k), x_{su}^m(k)) + 2 Y_{su}^f(k) (Y_{su}^f - Y_{su}^f(k)) \nonumber \\
     & + 2 x_{su}^m(k) (x_{su}^m - x_{su}^m(k)) . \ \ \label{eq:rate_affine_approx}
\end{align}
\noindent While the achievable rate in \eqref{eq: sbsOptimization} has been convexified, \eqref{eq: sbsOptimization} is still non-convex due to \eqref{eq:Lyapunov_Opt22}. By replacing the achievable rate in \eqref{eq:rate_affine} with \eqref{eq:Rate1Constraint}, the original constraint in \eqref{eq:Lyapunov_Opt22} is written as
\begin{align}
    \textstyle E_s(t) & \geq \frac {\transBits_{su}^f(t)} {\sbsTransDur \bwPerRB \realScheduling_{su}^f(t) \sum_{m=1}^{\rbsSize} x_{su}^m(t) + c}, \nonumber \\
    \textstyle \transBits_{su}^f(t) - & \sbsTransDur \bwPerRB E_s(t) \realScheduling_{su}^f(t) \sum_{m=1}^{\rbsSize} x_{su}^m(t) - c E_s(t) \leq 0. \nonumber
\end{align}
\indent Let $z_{su}^f(t) \geq 0$ be the new auxiliary variables such that:
\vspace{-0.3cm}
\begin{align}
\label{eq:Delay_constraint1}
 \textstyle \transBits_{su}^f(t) - c E_s(t) \leq E_s(t) z_{su}^f(t) \leq  \sbsTransDur \bwPerRB E_s(t) \realScheduling_{su}^f(t) \sum_{m=1}^{\rbsSize} x_{su}^m(t).
\end{align}
\indent By combining constraints in \eqref{eq:Rate1Constraint}, \eqref{eq:rate_affine} and \eqref{eq:Delay_constraint1}, the optimization problem \eqref{eq: Lyapunov_Opt2} is written as:
\vspace{-0.2cm}
\begin{subequations}
\label{eq: sbsOptimization2}
\begin{align}
    \underset{\boldsymbol{\delta}_{s}(t), \boldsymbol{\realScheduling_s}(t), \boldsymbol{E}(t), \boldsymbol{x}_s(t), \boldsymbol{z}_s(t)} {\text{maximize}} \ \ \ \textstyle \rbQueueEvo_{s}(t) \sum_{u \in \ueCoverage_s(t)} \sum_{m \in \rbsSize_s(t)} \delta_{su}^m(t) & - V E_s(t) + \sum_{u \in \ueCoverage_s(t)} \sum_{f=1}^{\contCatalog} \Bigg[ \virtQueue_{su}^f(t) \epsilon_u \bar{\packetArrRate}_{su}^f \qosReq_{s\ueRequests_uf}  \nonumber \\
      \textstyle + \frac {\sbsTransDur \bwPerRB} {2} ( \dataQueue_{su}^f(t) + \virtQueue_{su}^f(t) ) \Bigg\{ \underbrace{ \bigg( \realScheduling_{su}^f(t) + \Big[ \sum_{m=1}^{\rbsSize_s(t)} x_{su}^m(t) \Big) \Big] \bigg)^2 }_{g_2(\realScheduling_{su}^f, \realMatching_{su}^m)} & - \bigg( \underbrace{ \realScheduling_{su}^f(t)^2 + \Big[ \sum_{m=1}^{\rbsSize_s(t)} x_{su}^m(t) \Big]^2 }_{f_2(\realScheduling_{su}^f, \realMatching_{su}^m)} \bigg) \Bigg\} \Bigg], \ \label{eq:sbsOptimization21} \\
    \text{subject to} \ \ \textstyle z_{su}^f \leq \frac {\sbsTransDur \bwPerRB} {2} \bigg( g_2(\realScheduling_{su}^f(t), x_{su}^m(t)) - & f_2(\realScheduling_{su}^f(t), x_{su}^m(t)) \bigg), \label{eq:sbsOptimization22} \\
    \ \ \ \ \ \textstyle W_{su}^f + \underbrace{\frac {1} {2} \bigg( E_s(t)^2 + z_{su}^f(t)^2 \bigg) - c E_s(t)}_{f_3(z_{su}^f(t), E_{s}(t))} - & \underbrace{\frac {1} {2} \bigg( E_s(t) + z_{su}^f(t) \bigg)^2}_{g_3(z_{su}^f(t), E_{s}(t))} \leq 0, \label{eq:sbsOptimization26} \\
    \ \ \textstyle \sum_{u \in \ueCoverage_s} \sum_{m = 1}^{\rbsSize_s} \realMatching_{su}^m(t) \leq \rbQueueEvo_s(t), \ \ \ \ \ \ \ \ & \label{eq:sbsOptimization23} \\
    \ \ \textstyle \realMatching_{su}^m(t) \leq \sum_{f=1}^{\contCatalog} \realScheduling_{su}^f(t), \ \realScheduling_{su}^f(t) \in [0,1], &  \ \text{and} \ \eqref{eq: queue_eq1}, \eqref{eq: queue_eq2}, \eqref{eq:Rate1Constraint}, \ \ \forall s, u, m, \ \ \ \label{eq:sbsOptimization24}
\end{align}
\end{subequations}
where $\realScheduling_{su}^f(t)$ and $\realMatching_{su}^m(t)$ are the relaxed variables of $\schedVar_{su}^f(t)$ and $\matchPolicy_{su}^m(t)$. The first order approximation of $g_2(\realScheduling_{su}^f, x_{su}^m)$ in  \eqref{eq:sbsOptimization21}, \eqref{eq:sbsOptimization22}, and $g_3(E_{s}(t), z_{su}^f(t)$ in \eqref{eq:sbsOptimization26} respectively yields:
\vspace{-0.4cm}
\begin{align}
    \textstyle \hat{g}_2(\realScheduling_{su}^f,x_{su}^m;\realScheduling_{su}^f(k), x_{su}^m(k)) & = g_2(\realScheduling_{su}^f,x_{su}^m;\realScheduling_{su}^f(k), x_{su}^m(k)) + 2 \bigg( \realScheduling_{su}^f(k) + \sum_{m=1}^{\rbsSize_s(t)} x_{su}^m(k) \bigg) \bigg\{ (\realScheduling_{su}^f  \nonumber \\
    \textstyle & - \realScheduling_{su}^f(k)) + \Big( \sum_{m=1}^{\rbsSize_s(t)} x_{su}^m(k) \Big) \bigg\} \label{eq:Convexification_eq},
\end{align}
\vspace{-0.9cm}
\begin{align}
    \textstyle g_3(E_s, z_{su}^f;E_s(k), z_{su}^f(k)) = & g_3(E_s, z_{su}^f; E_s(k), z_{su}^f(k)) + ( E_s(k) + z_{su}^f(k) ) \big\{ (E_s - E_s(k)) + \nonumber \\
    & (z_{su}^f - z_{su}^f(k)) \big\}. \label{eq:Convexification_eq2}
\end{align}
\indent Since all the terms in \eqref{eq: sbsOptimization2} are either convex or a difference of convex functions, it can be solved by invoking the CCP procedure \cite{Ref1:CCP}. 
\footnotesize
\LinesNumberedHidden{
\begin{algorithm}[t]
\caption{CCP Algorithm for solving \eqref{eq: sbsOptimization2}}
\label{algo:algo2}
\DontPrintSemicolon 
\While{$s \in \sbsSet$} {
    Initialize $i = 0$, and a feasible point $(E_s^{i},\realMatching_{su}^{(m,i)}, \realScheduling_{su}^{(f,i)})$ in \eqref{eq:sbsOptimization21}, \eqref{eq:sbsOptimization23}, \eqref{eq:sbsOptimization24} \;
    \Repeat{convergence} {
        Convexify $g_2(\realScheduling_{su}^f, x_{su}^m)$ and $g_3(E_{s}(t), z_{su}^f(t)$ in \eqref{eq:sbsOptimization21} and \eqref{eq:sbsOptimization26} according to \eqref{eq:Convexification_eq} \;
        Using \eqref{eq:Convexification_eq} and \eqref{eq:Convexification_eq2} for $g_2(\realScheduling_{su}^f, x_{su}^m)$ and $g_3(E_{s}, z_{su}^f)$, solve the convex problem \eqref{eq: sbsOptimization2} \;
        Find the optimal $E_s^{(i)*},\realMatching_{su}^{(m,i)*}, \realScheduling_{su}^{(f,i)*}$ \;
        Update $E_s^{(i+1)} := E_s^{(i)*}, \realMatching_{su}^{(m,i+1)} := \realMatching_{su}^{(m,i)*}, \realScheduling_{su}^{(f,i+1)} := \realScheduling_{su}^{(f,i)*}$  and $i := i+1$ \;
    }
}
\end{algorithm}}
\normalsize
\subsubsection{Fronthaul allocation at the cloud}
From \eqref{eq:Cluster_Utility}, \eqref{eq:sbss_utility}, \eqref{eq:cloud_utility}, the fronthaul allocation problem is independent of the matching policy problem. Therefore, the fronthaul allocation problem at the cloud aims to reduce the cost at the cloud. The main goal is to minimize \eqref{eq:opt_eq} for the above problem. Therefore, the problem is given as:
\vspace{-0.3cm}
\begin{subequations}
\label{eq:frthaulOpt}
\begin{align}
    \underset{\textstyle \boldsymbol{\frthaulPerUe}(t), B(t)}{\text{maximize}} \ & \ \textstyle \sum_{s=1}^{\sbsSize} \sum_{f=1}^{\contCatalog} \Big[ (1-\sbsTransDur) \dataQueue_{cs}^f(t) \textstyle \frthaulPerUe_{s}^f \frthaulCap + \virtQueue_{cs}^f(t) \Big( \textstyle (1-\sbsTransDur) \frthaulPerUe_{s}^f \frthaulCap - \frac {\epsilon_s \qosReq_{s\ueRequests_uf}} {\packetSize_f} \Big) \Big] - V B(t), \ \label{eq:frthaulOpt1} \\
    \text{subject to} \ \ & \textstyle \sum_{s \in \sbsSet} \boldsymbol{1}\IEEEauthorrefmark{2} \textstyle \boldsymbol{\frthaulPerUe}_s \leq 1, \ \text{and} \ \eqref{eq: queue_eq3}, \eqref{eq:Obj_Func8}, \eqref{eq:Lyapunov_Opt23}, \eqref{eq:Deficit_queue2}. \label{eq:frthaulOp_2} 
\end{align}
\end{subequations}
\indent The optimization problem is non-convex due to \eqref{eq:Lyapunov_Opt23} and belongs to DC programming problem. By using the mathematical rule $xy = (x+y)^2 - x^2 - y^2$ and rewriting $\eqref{eq:Lyapunov_Opt23}$ yields:
\vspace{-0.3cm}
\begin{align*}
    \textstyle \transBits_{cs}^f(t) & \leq \underbrace{\big\{ B(t) + (1-\sbsTransDur) \frthaulPerUe_{s}^f \frthaulCap + c \big\}^2}_{f_4(x)} - \underbrace{ \big\{ B^2(t) + \big\{ (1-\sbsTransDur) \frthaulPerUe_{s}^f \frthaulCap + c \big\}^2 \big\}}_{g_4(x)}.
\end{align*}
\indent Problem \eqref{eq:frthaulOpt} is solved using the same CCP procedure defined in Algorithm~\ref{algo:algo2} which results in $\frthaulPerUe_s^{(f)*}, B^*$. The gradient of $g_0(x)$
\vspace{-0.3cm}
\begin{align*}
    \textstyle \hat{g}_i(B, \frthaulPerUe; B_k, \frthaulPerUe_k) & = g_i(B_k, \frthaulPerUe_k) + \nabla^{\IEEEauthorrefmark{2}}g_i(B_k, \frthaulPerUe_k) \big(B - B_k; \frthaulPerUe - \frthaulPerUe_k\big). \nonumber \\
    \textstyle \nabla g(B, \frthaulPerUe) & = \big(2 B; 2[ (1-\sbsTransDur) \frthaulPerUe_{s}^f \frthaulCap + c] (1-\sbsTransDur)\frthaulCap\big). \nonumber
\end{align*}

\vspace{-0.9cm}
\subsection{Convergence of Alg. \ref{algo:algo2}}
\vspace{-0.2cm}
To prove convergence of resource allocation and scheduling problem, we will focus on the convergence of \eqref{eq:sbsOptimization22}. The convergence of \eqref{eq:sbsOptimization26} follows similar steps. Consider the simplified form of \eqref{eq:sbsOptimization22} in the CCP at iteration $k$:
\vspace{-0.2cm}
\begin{align*}
    \textstyle P_k & = \underbrace{\Big(\realScheduling_{su}^f(k) + \sum_{m=1}^{\rbsSize_s(t)} x_{su}^m(k) \Big)^2}_{g_2(\realScheduling_{su}^f(k), x_{su}^m(k))} - \underbrace{\bigg(\realScheduling_{su}^f(k)^2 + \Big(\sum_{m=1}^{\rbsSize_s(k)} x_{su}^m(k)\Big)^2 \bigg)}_{f_2(\realScheduling_{su}^f(k), x_{su}^m(k))}, \nonumber \\
    \textstyle & = \underbrace{\Big(\realScheduling_{su}^f(k) + \sum_{m=1}^{\rbsSize_s(t)} x_{su}^m(k) \Big)^2}_{\hat{g}_2(\realScheduling_{su}^f(k), x_{su}^m(k); \realScheduling_{su}^f(k), x_{su}^m(k))} - \underbrace{\bigg(\realScheduling_{su}^f(k)^2 + \Big(\sum_{m=1}^{\rbsSize_s(k)} x_{su}^m(k)\Big)^2 \bigg)}_{f_2(\realScheduling_{su}^f(k), x_{su}^m(k))}, \nonumber \\
    \textstyle & \leq \underbrace{\Big(\realScheduling_{su}^f(k) + \sum_{m=1}^{\rbsSize_s(t)} x_{su}^m(k) \Big)^2 + 2 \Big(\realScheduling_{su}^f(k) + \sum_{m=1}^{\rbsSize_s(t)} x_{su}^m(k) \Big) \Big\{ (\realScheduling_{su}^f(k+1) - \realScheduling_{su}^f(k)) }_{\hat{g}_2(\realScheduling_{su}^f(k+1), x_{su}^m(k+1); \realScheduling_{su}^f(k), x_{su}^m(k))} \nonumber \\
    & \ \ \ \textstyle + \underbrace{\bigg(\sum_{m=1}^{\rbsSize_s(k+1)} x_{su}^m(k+1) - \sum_{m=1}^{\rbsSize_s(k)} x_{su}^m(k)\bigg) \Big\}}_{\hat{g}_2(\realScheduling_{su}^f(k+1), x_{su}^m(k+1); \realScheduling_{su}^f(k), x_{su}^m(k))} - f_2(\realScheduling_{su}^f(k+1), x_{su}^m(k+1)),
\end{align*}
\indent where the last inequality is due to the fact that $z_{su}^f > 0$ which corresponds to maximizing the CCP approximation in \eqref{eq:sbsOptimization22}. By choosing $P_{k+1} = P_k$, we maximize the value of $f_2(\realScheduling_{su}^f(k), x_{su}^m(k)) - \hat{g}_2(\realScheduling_{su}^f, x_{su}^m; \realScheduling_{su}^f(k), x_{su}^m(k))$. Therefore
\begin{equation*}
    \textstyle P_k \leq f_0(\realScheduling_{su}^f(k+1), x_{su}^m(k)) - \hat{g}_0(\realScheduling_{su}^f(k+1), x_{su}^m(k+1); \realScheduling_{su}^f(k), \textstyle x_{su}^m(k)) \leq P_{k+1}.
\end{equation*}
\indent Therefore the sequence $\{P_k\}_{i=0}^{\infty}$ is non decreasing and convergent.
\vspace{-0.4cm}
\section{Simulations}
\begin{figure}[b!]%
\centering
\captionsetup{justification=centering}
\vspace{-0.3cm}
\includegraphics[scale = 0.33]{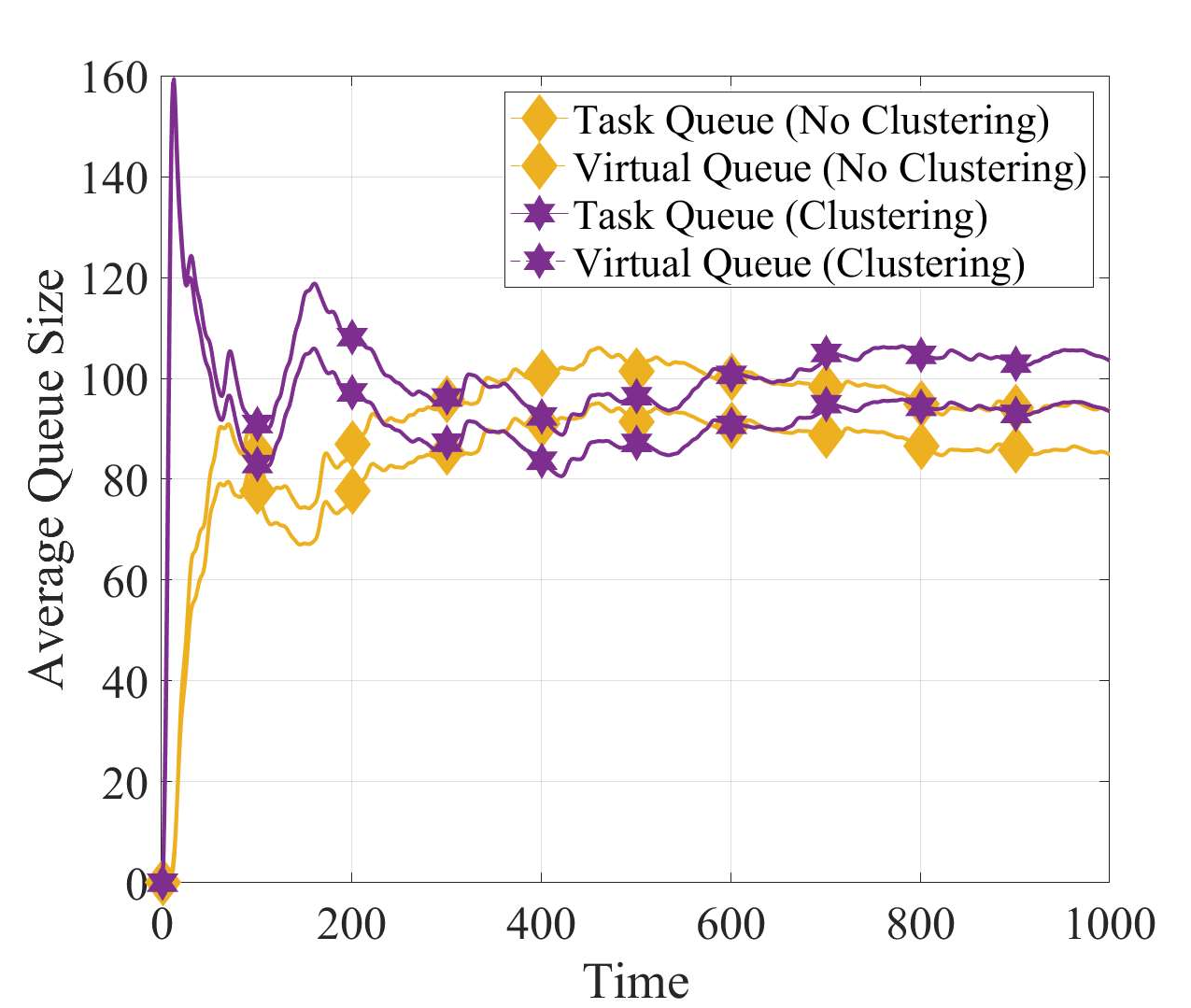}
\vspace{-0.5cm}
\caption{Convergence of the proposed schemes.}
\label{fig:Convergence}
\end{figure}

\begin{figure}[t]
    \centering
    \captionsetup{justification=centering}
    \begin{minipage}{.5\textwidth}
        \centering
        \vspace{-1.5cm}
        \includegraphics[scale = 0.29]{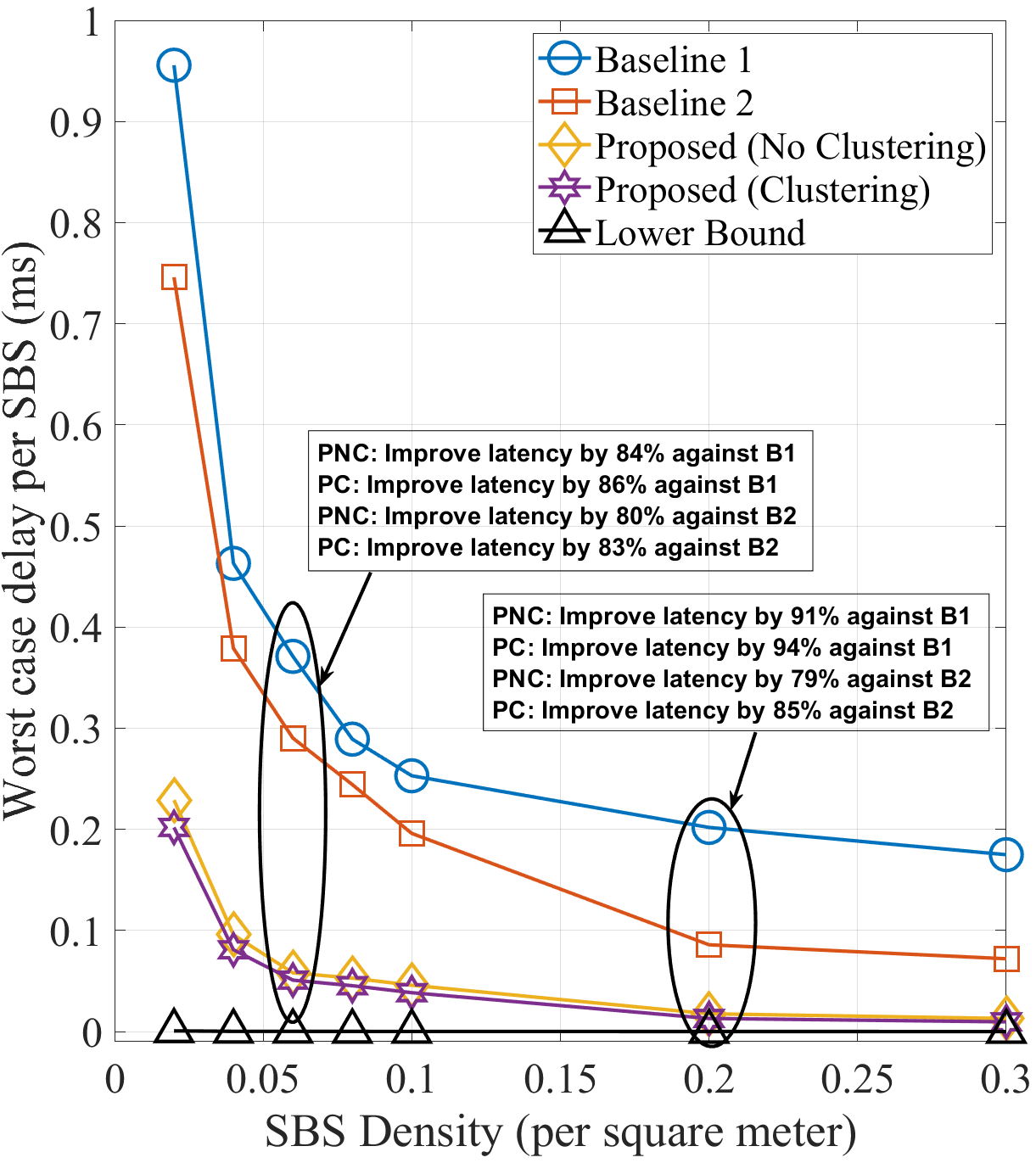}
        \vspace{-1.0cm}
        \caption{Comparison of the worst case delay for $\ueDensity = 0.04$}
        \label{fig:sbsDensity1}
    \end{minipage}%
    \begin{minipage}{0.5\textwidth}
        \centering
        \vspace{-1.5cm}
        \includegraphics[scale = 0.295]{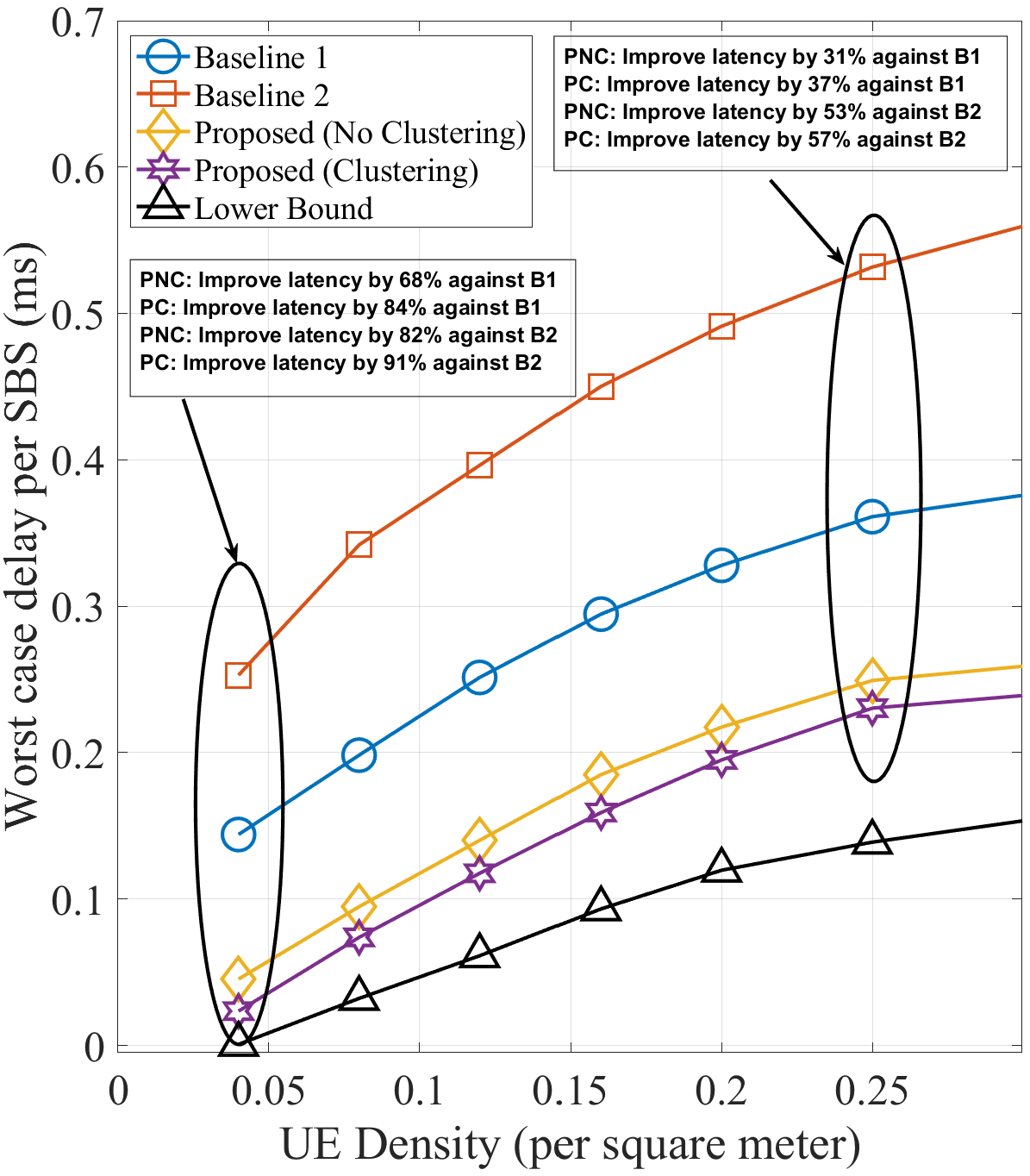}
        \vspace{-1.1cm}
        \caption{Comparison of the worst case delays for $\sbsDensity = 0.02$}
        \label{fig:ueDensity}
    \end{minipage}
    \vspace{-0.5cm}
\end{figure}

\vspace{-0.1cm}
We consider an area of $10\mathrm{m} \times 10\mathrm{m}$ where SBSs and UEs are distributed uniformly according to $\sbsDensity$ and $\ \ueDensity$. The total transmission power of SBSs is $23$dBm. Other simulation parameters are listed in Table. \ref{table:SimParam}. To evaluate the performance of the proposed schemes with clustering (PC) and proposed method with No-clustering (PNC), two baseline methods are considered: \textit{Baseline 1} (B1) and \textit{Baseline 2} (B2). The details of both baselines are shown in Table \ref{table:Simulation_Methods}. In addition, we showed the upper bound of the performance in Fig. \ref{fig:sbsDensity1} and Fig. \ref{fig:ueDensity} as the lower bound of the delay. \\
\indent Fig. \ref{fig:Convergence} shows the convergence of the proposed methods. In the beginning, the queue size for PC increases instantly. However, it starts converging at $t = 300$ in which a gradual learning can be seen that achieves convergence around t = 500. The reason for the slow variation in PNC is that the estimated interference learnt in \eqref{eq:EstimatedInterference1} slowly converges to a steady state due to increased intra-SBS interference. With PC, SBSs experience less interference from other SBSs due to clustering. As a result, the estimated interference converges faster than PNC.
\begin{table}[t]
\begin{minipage}{.5\linewidth}
    \centering
    \caption{Simulation Parameters}
    \vspace{-0.3cm}
    \label{table:SimParam}
    \begin{tabular}{ p{3.2cm} c c } 
        \toprule
        \makecell{Parameter} & Symbol & Value \\   
        \midrule
        LibrarySize & $\contCatalog$ & 20 \\
    	Bandwidth & $\bwPerRB$ & $1.4$ MHz \\ %
    	SBS Clustering time & $\sbsClustTime$ & 300 \\
    	Cache update time & $\cacheUpdateTime$ & 10 \\
    	User preference time & $\uePrefTime$ & 10 slots \\
    	SBSs/cloud temperature coefficient & $\CacheTempCoeff_s/\CacheTempCoeff_c$ & 0.02/0.05 \\
    	RL learning parameters & $\learParam_i$ & \{0.55,0.65,0.75\} \\
    	Noise power & $\noise$ & $-174$dBm/Hz \\
    	Packet Size & & 2000 bytes\cite{Chapter1_MatchingTheory5} \\
        \bottomrule
    \end{tabular}
\end{minipage}\hfill
\begin{minipage}{.5\linewidth}
    \centering
    \caption{Simulation Methods}
    \vspace{-0.5cm}
    \label{table:Simulation_Methods}
    \medskip
    \begin{tabular}{ l p{2.5cm} p{2.5cm} } 
        \toprule
        \makecell{} & \textbf{Baseline 1 (B1)} & \textbf{Baseline 2 (B2)} \\   
        \midrule
        User Association & Random  & Nearest cached content \\
        RB Allocation & Random served on an RB & Received Signal Strength Indicator (RSSI)-based \\
        User scheduling & Random &  Proportional fair \\
        Content popularity & Same popularity of contents i.e., uniform & Time-varying \\
        Caching strategy & Random & Average content requests frequency \\
        \bottomrule
    \end{tabular}
\end{minipage}
\end{table}

\begin{figure}[b]
    \centering
    \captionsetup{justification=centering}
    \begin{minipage}{.5\textwidth}
        \centering
        \vspace{-0.3cm}
        \includegraphics[scale=0.32]{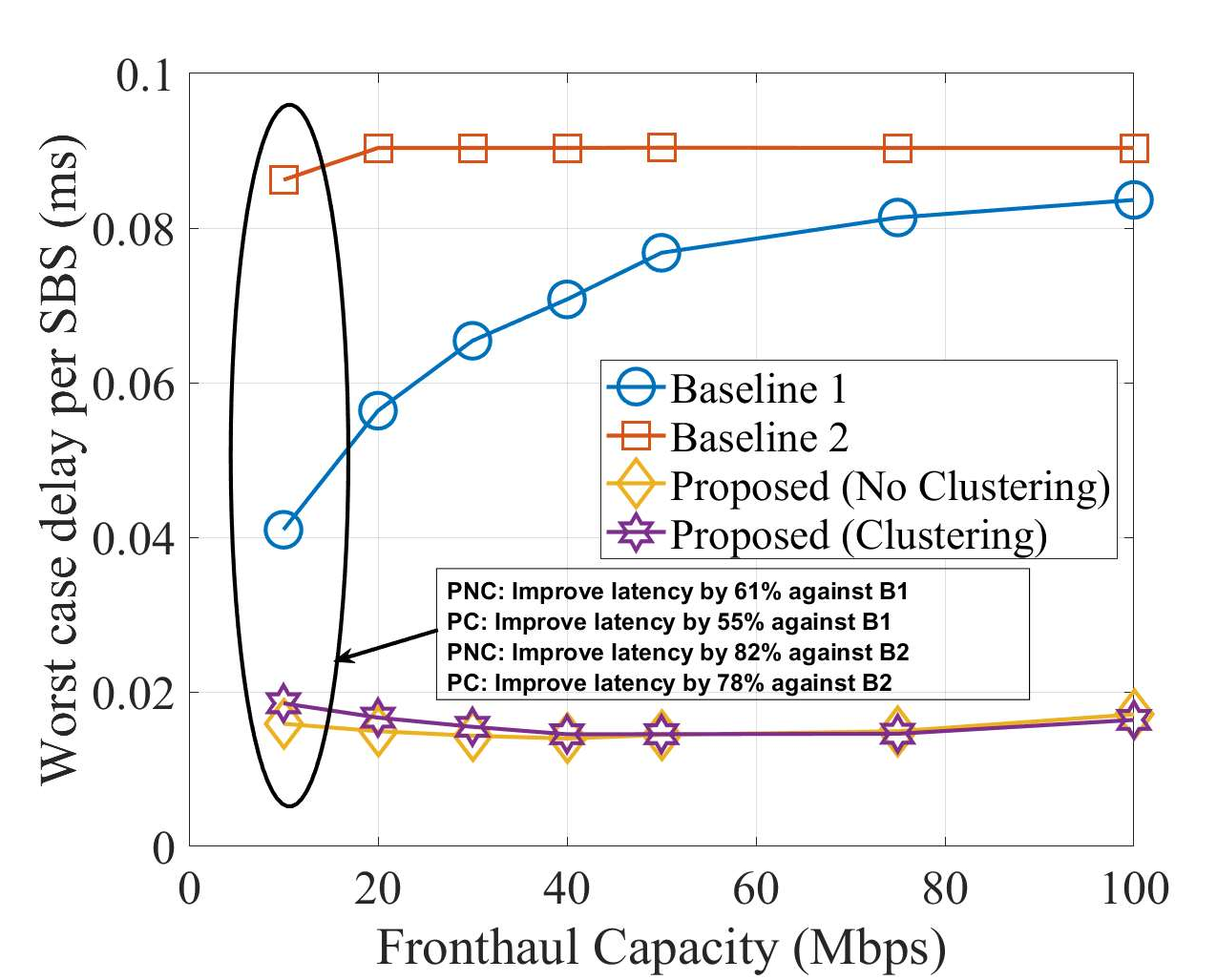}
        \vspace{-0.5cm}
        \caption{Comparison of worst case delays for $\ueDensity = 0.04$, $\sbsDensity = 0.06$ for different values of fronthaul capacity}
        \label{fig:Fronthaul}
    \end{minipage}%
    \begin{minipage}{0.5\textwidth}
        \centering
        \vspace{-0.3cm}
        \includegraphics[scale=0.32]{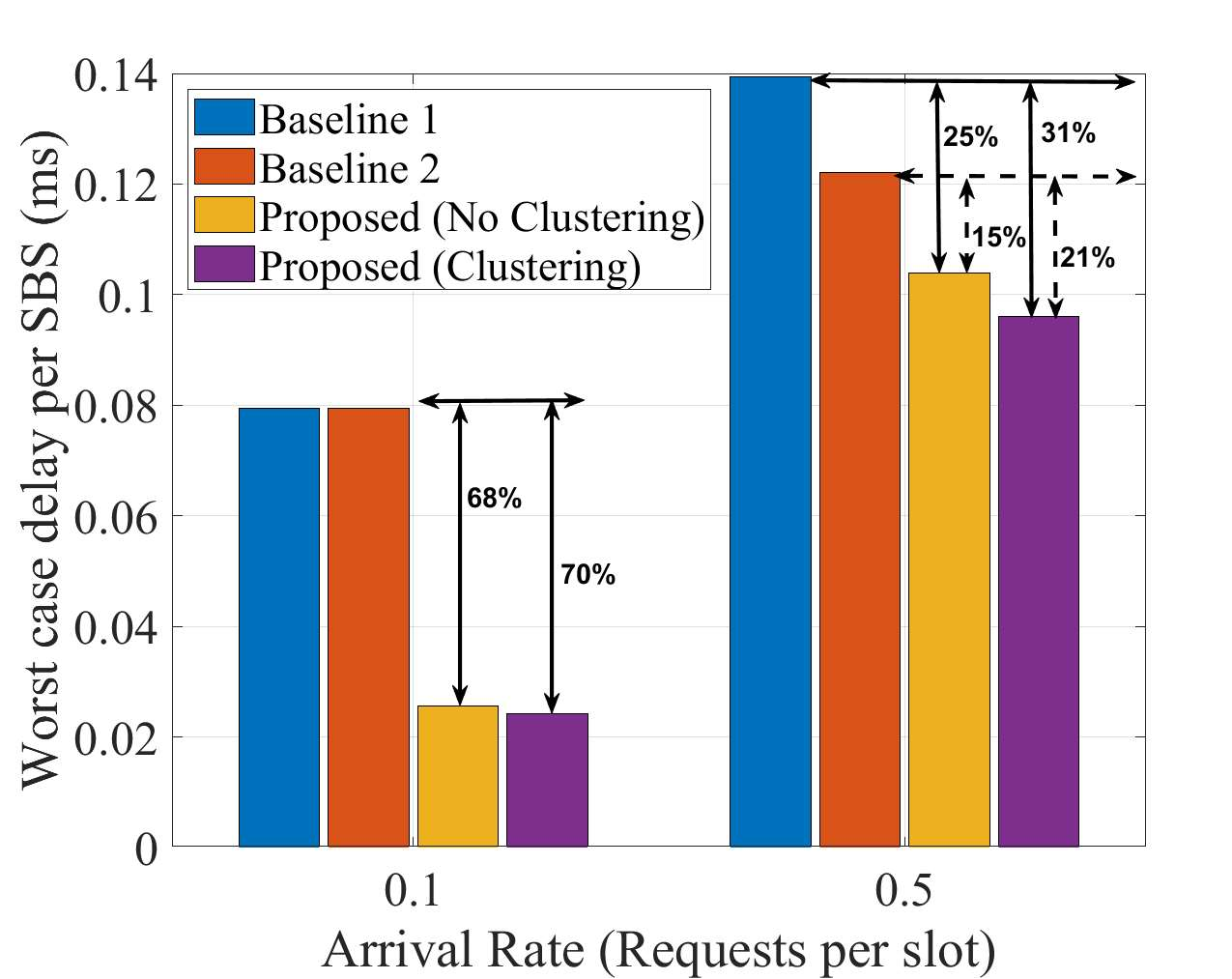}
        \vspace{-0.5cm}
        \caption{Comparison of worst case delays for different values of arrival rate for $\ueDensity = 0.04$, $\sbsDensity = 0.06$}
        \label{fig:Arrival}
    \end{minipage}
    \vspace{-0.5cm}
\end{figure}

\indent Fig. \ref{fig:sbsDensity1} shows the worst case delay as a function of SBS Density. With the increased SBS density for a fixed number of users, the worst-case packet delay, calculated by aggregating $\textstyle \bar{\aggCost}_{s}(\boldsymbol{\insCache}_s^{\sbsClustTime}, \boldsymbol{\mathrm{\matchPolicy}}_s^{\cacheUpdateTime})$ and $\textstyle \bar{\aggCost}_{c}(\boldsymbol{\insCache}_s^{\sbsClustTime}, \boldsymbol{\frthaulPerUe}^{T_2})$, decreases due to the increased cumulative cache size. The total number of contents jointly cached by the SBSs increases the probability of cache hits. Depending on the number of users and arrival rate, the increased number of SBSs increases the interference among SBSs yielding low rates and high delays. At the same time, the proposed approach reduces the delay compared to both baselines. 
Specifically, Fig \ref{fig:ueDensity} shows the average delay of the proposed approach as a function of UE Density. With increased UE density for a fixed number of SBSs, the worst-case packet delay increases due to the increased number of requests resulting in increased queue backlog at the SBSs. Hence, all SBSs attempt to schedule multiple users simultaneously causing high interference and thus, higher delays.
\begin{figure}[t]
    \centering
    \begin{minipage}{.5\textwidth}
        \centering
        \captionsetup{justification=centering}
        \vspace{-0.3cm}
        \includegraphics[scale=0.32]{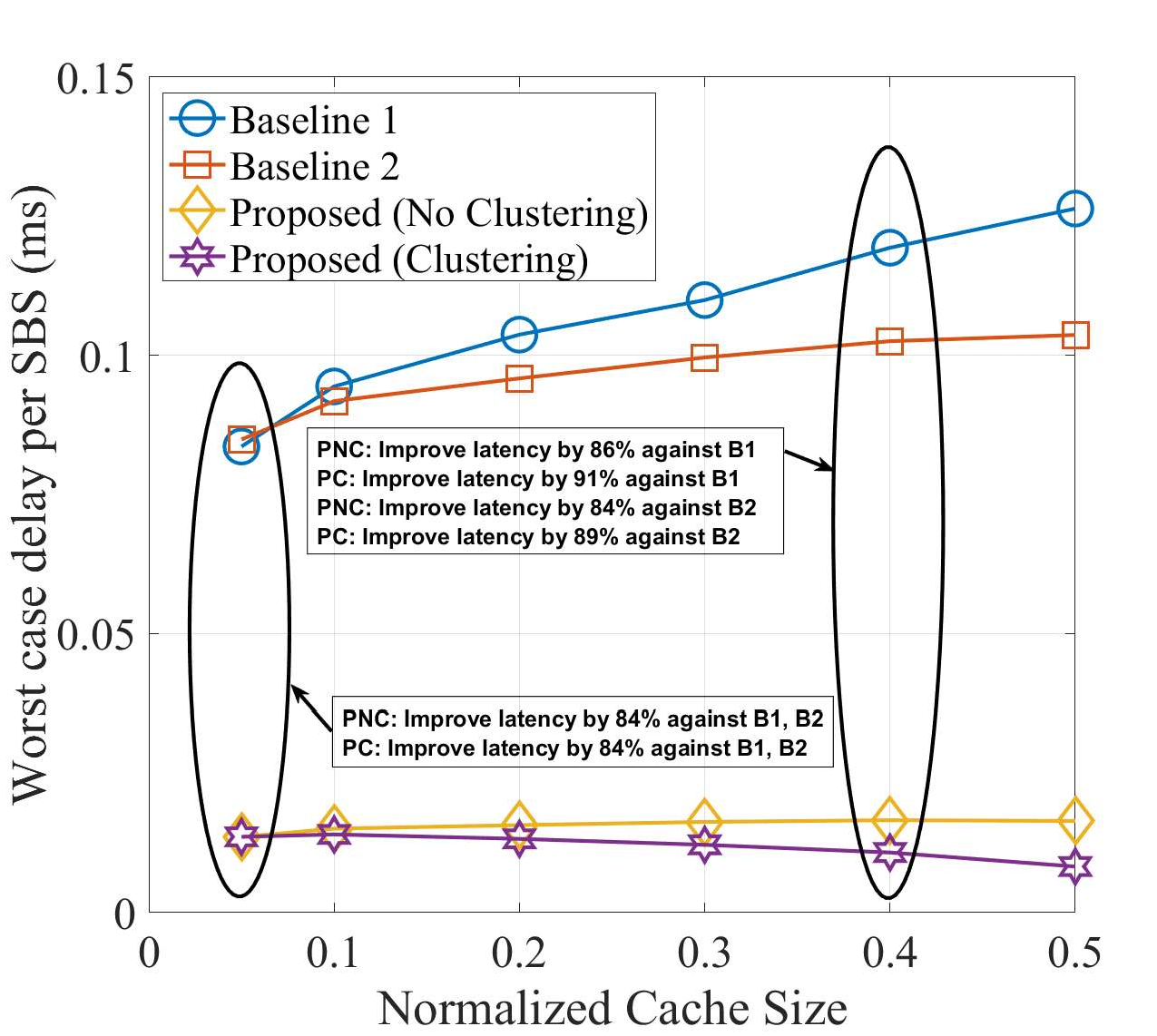}
        \vspace{-0.5cm}
        \caption{Comparison of proposed schemes for different cache size}
        \vspace{-0.3cm}
        \label{fig:CacheSize}
    \end{minipage}%
    \begin{minipage}{0.5\textwidth}
        \centering
        \captionsetup{justification=centering}
        \vspace{-0.15cm}
        \includegraphics[width=6.5cm, height=6.1cm]{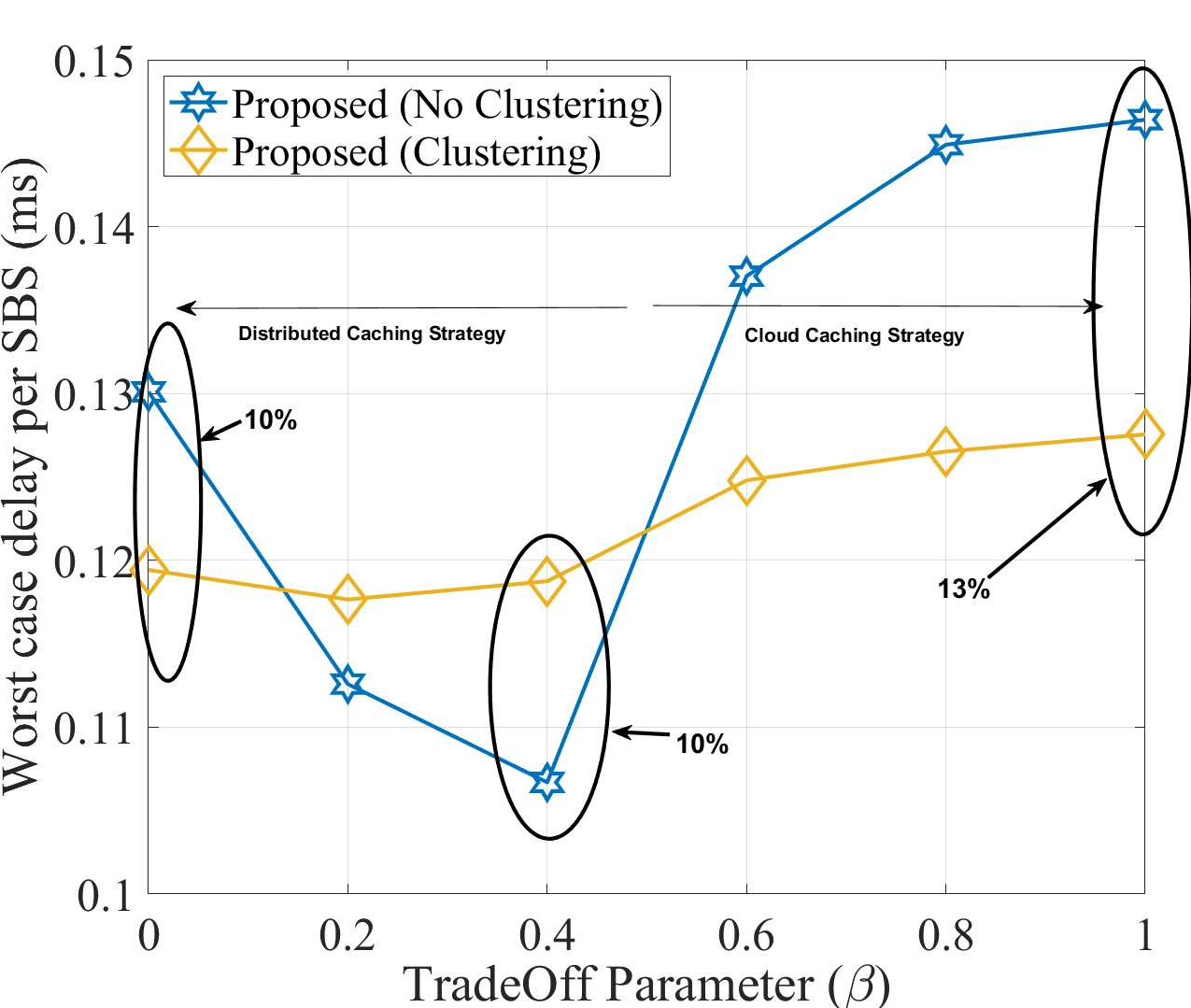}
        \vspace{-0.5cm}
        \caption{Impact of Cloud aided caching strategy with distributed caching strategy}
        \vspace{-0.3cm}
        \label{fig:TradeOff}
    \end{minipage}
\end{figure}

\indent Fig \ref{fig:Fronthaul} shows the worst-case delay of the proposed approach as a function of fronthaul capacity. The figure shows that the delay increases with the increased fronthaul capacity in B1 due to the randomness in caching strategy, user scheduling and resource allocation. With the increased fronthaul capacity, the total packet delay of the proposed approach remains the same for a fixed UE density and SBS density. For a fixed number of users and SBSs, the delay incurred in transmitting content to the user over the access link remain the same for a fixed arrival rate. However, the delay improves slightly over the fronthaul link. Fig. \ref{fig:Arrival} shows the worst-case delay of the proposed approach for two different values of arrival rate. With the increased arrival rate for a fixed number of users and SBSs, the worst-case packet delay increases due to the increased requests of user. The increased arrival rate increases the queue backlog at SBSs attempting to schedule more users resulting in increased interference thereby increasing the packet delay. The figure also shows the benefit of clustering for a high arrival rate.
\begin{figure}[b!]
    \centering
    \begin{minipage}{.5\textwidth}
        \centering
        \captionsetup{justification=centering}
        \vspace{-1.7cm}
        \includegraphics[scale=0.32]{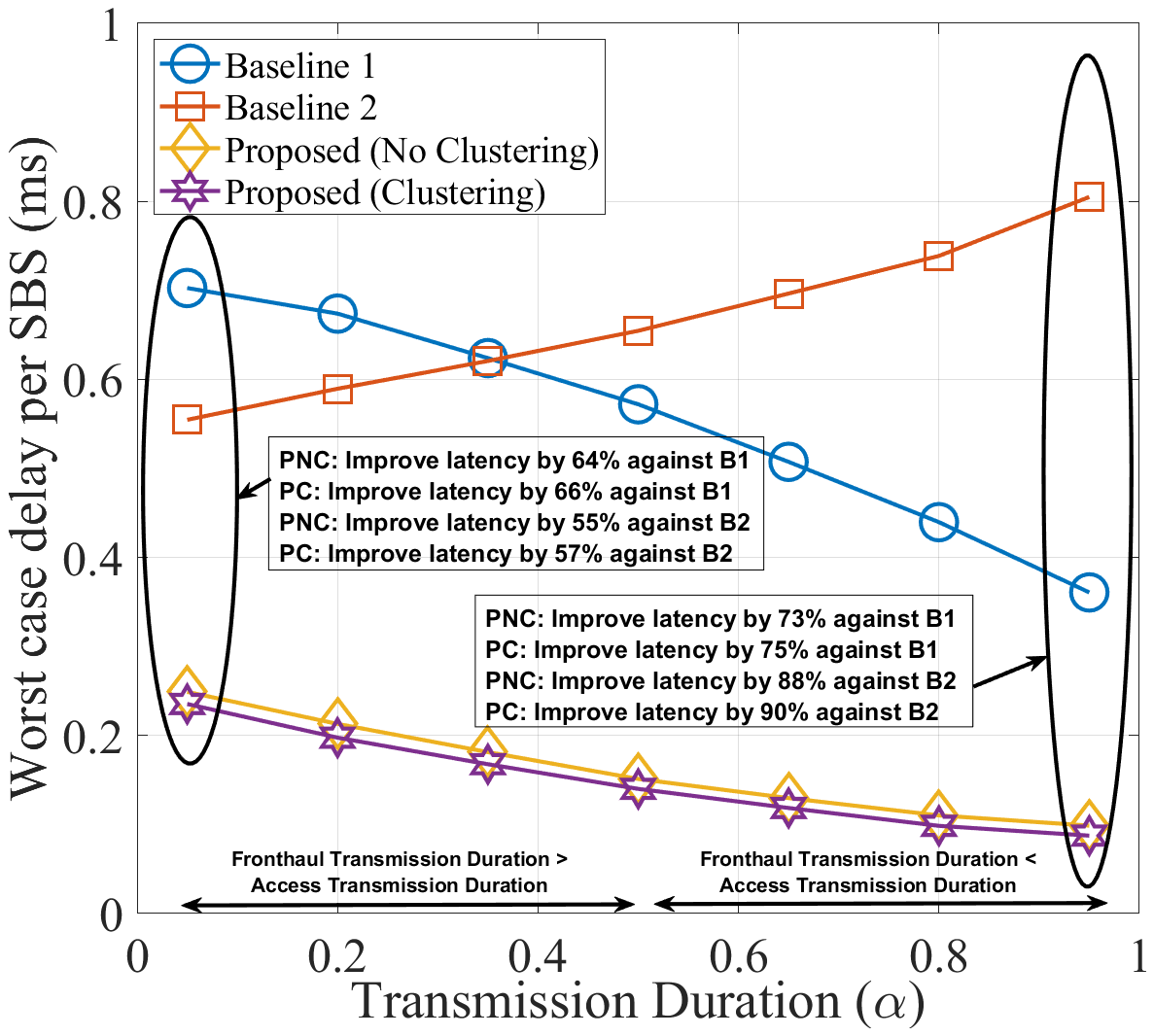}
        \vspace{-1.9cm}
        \caption{Comparison of the proposed scheme with baseline for different values of transmission duration for $\ueDensity = 0.04$, $\sbsDensity = 0.06$}
        \label{fig:TransDuration}
    \end{minipage}%
    \begin{minipage}{0.5\textwidth}
        \centering
        \captionsetup{justification=centering}
        \vspace{-2.6cm}
        \includegraphics[width=7cm, height=9.25cm]{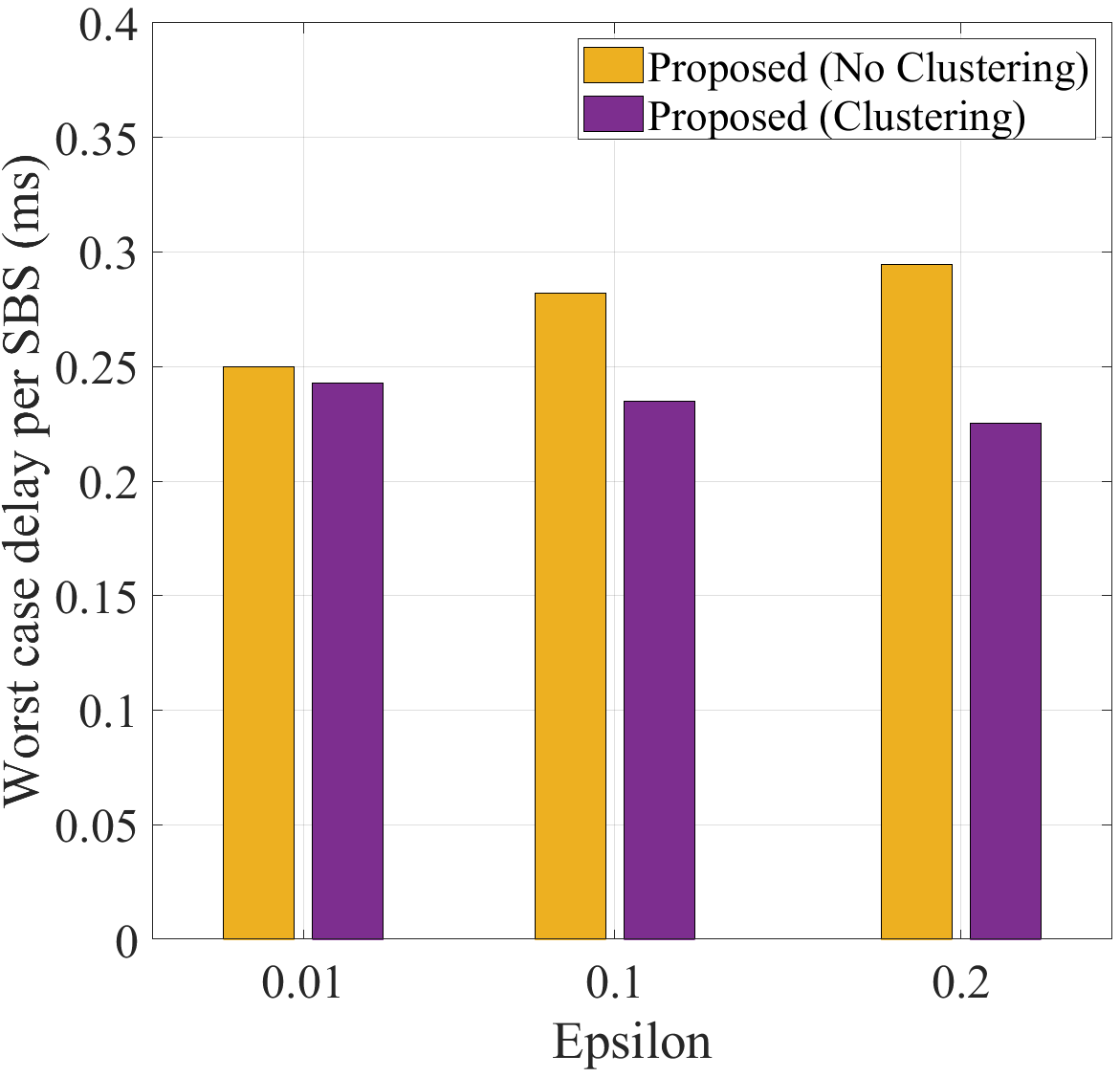}
        \vspace{-2.2cm}
        \caption{Impact of latency constraints on the proposed schemes for $\ueDensity = 0.04$, $\sbsDensity = 0.06$}
        \vspace{0.6cm}
        \label{fig:Epsilon}
    \end{minipage}
\end{figure}

\indent Fig. \ref{fig:CacheSize} shows the worst-case delay for different cache size. With the increased cache size, cache hit probability increases. As the SBSs serve the user requests instantly. Serving multiple users increases the interference resulting in increases packet delay. Due to the ability to reduce queuing and possibility of increasing interference are having a tradeoff, the worst-case packet delay for the proposed approaches shows little change with increased cache size. In addition, B1 experiences increased delay with cache size due to the random resource allocation. However, B2 shows slightly better performance due to RSSI based resource allocation. Fig. \ref{fig:TradeOff} shows the effect of the local/global tradeoff parameter on the proposed approach. When the SBSs take caching and resource allocation decisions based entirely on their local/global environment, PC will improve the delay due to the reduced interference from neighboring SBSs. On the other hand, caching and resource allocation based on a mixture of local and global decisions degrade the performance of PC. The reason for such behavior is the excess of content requests from users not in the vicinity of SBSs as well as the excess interference from other SBSs. \\
\indent Fig. \ref{fig:TransDuration} shows the worst-case delay of the proposed strategy as a function of transmission duration parameter $\sbsTransDur$. With a small value of transmission duration, the SBSs could not have enough time to serve user requests as per \eqref{eq:basicRate}. This increases the queue backlog at SBSs resulting in increased delay. For a large value of transmission duration, the content transmission over the fronthaul link for a content not cached by the SBSs decreases which does not increase the average queue size at the SBSs causing a low delay. While B2 shows smaller delay than B1 for a small value of transmission duration. There is a cutoff where B2 shows larger delay than B1 due to increased interference than B1. While B1 assumes random user association, B2 considers nearest neighbor association which results in more severe interference than random user association. \\ 
\indent Fig. \ref{fig:Epsilon} shows the effect of latency constraints \eqref{eq:Obj_Func3} and \eqref{eq:Obj_Func4} of the proposed approach. The figure shows the proposed approach without clustering is highly sensitive to latency requirements. For a very stringent latency requirements, the worst-case packet delay for both proposed methods is the same. However, with the relaxed latency requirements, the proposed approach without clustering shows an increased value of delay. On the other hand, the proposed approach with clustering shows the delay to be slightly decreased for different values of latency requirements.
\vspace{-0.4cm}
\section{Conclusion}
\vspace{-0.1cm}
In this paper, we have proposed a new scheme for caching, resource allocation and scheduling in cloud-aided wireless networks while satisfying queueing latency constraints. To solve the problem, we have introduced a two-timescale optimization strategy : SBSs learn caching strategy based on the spatio-temporal user behavior over a longer timescale while a joint scheduling and dynamic matching based resource allocation algorithm serves the requests of the users at each time slot. Simulation results have shown that our proposed scheme outperforms baseline approaches and achieves gains of up to 90\%  in terms of packet delay.
\vspace{-0.5cm}
\input{Appendix.tex}

\end{document}

%% file: Appendix.tex
\appendices
\section{Proof of Convergence of RL based caching strategy}
\label{sec: Proof_Lemma2}
Let us assume that $x_{su}(t) > y_{su}(t) > 0$ at time $t$ and consider $\textstyle E_u(t) = \abs{\frac {1} {\boldsymbol{\cost_{su}}}} = $\\
\noindent $\abs{ \frac{\log_2 (1 + x_{su}(t)) - \log_2 (1 + y_{su}(t))} {x_{su}(t) - y_{su}(t)}}$ which is reordered as follows;
\begin{align*}
    \textstyle E_u(t) & = \textstyle \abs{ \textstyle \frac{\textstyle \log_2 (1 + \textstyle x_{su}(t)) - \textstyle \log_2 (1 + \textstyle y_{su}(t))} {x_{su}(t) - \textstyle y_{su}(t)}} \nonumber \\
    & = \textstyle \abs{ \frac {1} {1 + \textstyle y_{su}(t)} \frac {\textstyle \textstyle \ln{(1 + \textstyle z_{su}(t))}} {z_{su}(t) \ln{2}}} \leq \textstyle \frac { \textstyle \ln{(1 + \textstyle z_{su}(t))}} {\textstyle z_{su}(t) \ln{2}}
\end{align*}
where $z_{su}(t) = \frac {x_{su}(t) - y_{su}(t)} {1 + y_{su}(t)}$. To find the upper bound, consider the first derivative of $\frac {\ln{(1 + z_{su}(t))}} {z_{su}(t)}$:
\begin{equation*}
    \textstyle \frac {\textstyle d} {\textstyle dz_{su}(t)}\Big( \textstyle \frac {1 + \textstyle \ln{(1 + \textstyle z_{su}(t))}} {\textstyle z_{su}(t)} \Big) = \textstyle \frac {\textstyle 1} {\textstyle z_{su}^2(t)} \textstyle \Big(1 - \textstyle \frac {\textstyle 1} {\textstyle 1 + z_{su}(t)} - \textstyle \ln{(1 + z_{su}(t))} \Big) < \textstyle 0; \forall  z_{su}(t) \geq 0. 
\end{equation*}
This yields that the upper bound of $\frac {\ln{(1 + z_{su}(t))}} {z_{su}(t)}$ is $\lim_{z \to 0} \frac {\ln{(1 + z_{su}(t))}} {z_{su}(t)} = 1$. Therefore $E_u(t) \leq \log_{2} e$ and $\log_2 (1 + x_{su}(t)) - \log_2 (1 + y_{su}(t)) \leq L(x_{su}(t) - y_{su}(t))$ for a scalar $L$, i.e., $\tilde{\utility}_{s}$ is Lipschitz \cite{Convergence1}. With the satisfactory of above conditions, using \cite[Equation. 7 and Proposition 4.1]{Convergence2} and the fact that $c$ is a constant, the convergence is achieved.
\vspace{-0.5cm}
\section{Proof of the Lemma 1}
\vspace{-0.2cm}
\label{sec: Proof_Lemma1}
Squaring the queue dynamic equation \eqref{eq: queue_eq1} and \eqref{eq: queue_eq2}, we have:
\begin{align}
\textstyle \dataQueue_{su}^{f}(t+1)^2 & \ = \Big(\mathrm{max}[\dataQueue_{su}^f(t) - \sum_{m=1}^{\rbsSize}\rate_{su}^{(m,f)}(t), 0] + \packetArrRate_{su}^f(t)\Big)^2 \leq \dataQueue_{su}^{f}(t)^2 + \packetArrRate_{su}^{f}(t)^2 \nonumber \\
& \textstyle - 2 \dataQueue_{su}^f(t)\Big(\sum_{m=1}^{\rbsSize}\rate_{su}^{(m,f)}(t) - \packetArrRate_{su}^f(t)\Big) + \Big(\sum_{m=1}^{\rbsSize}\rate_{su}^{(m,f)}(t)\Big)^2 \label{eq:lemma11} \\
\textstyle \dataQueue_{cs}^f(t+1)^2 & = \Big( \mathrm{max}[\dataQueue_{cs}^f(t) - (1-\sbsTransDur) \frthaulPerUe_{s}^f \frthaulCap ,0] + \frac {1} {\packetSize_f} \sum_{u=1}^{\ueSize} \mathbbm{1}_{\{q_u(t) \not\in \insCache_s(t) \} } \Big)^2 \leq \dataQueue_{su}^{f}(t)^2  + \Big((1-\sbsTransDur) \frthaulPerUe_{s}^f\Big)^2 \nonumber \\
 \textstyle - 2 \dataQueue_{su}^f(t)\Big( & (1-\sbsTransDur) \frthaulPerUe_{s}^f \frthaulCap - \frac {1} {\packetSize_f} \sum_{u \in \ueCoverage_s(t)} \mathbbm{1}_{\{q_u(t) \not\in \insCache_s(t) \} } \Big) + \Big( \frac {1} {\packetSize_f} \sum_{u \in \ueCoverage_s(t)} \mathbbm{1}_{\{q_u(t) \not\in \insCache_s(t) \}} \Big)^2 \label{eq:lemma14} \\
\textstyle \rbQueueEvo_{s}^2(t+1) & \ \leq \ \rbQueueEvo_{s}^2(t) - 2 \rbQueueEvo_{s}(t)\bigg(\sum_{u \in s(u)} |\boldsymbol{\matchPolicy}_{su}(t)| - \rbsArrRate_{s}(t)\bigg) + \bigg(\sum_{u \in s(u)} |\boldsymbol{\matchPolicy}_{su}(t)|\bigg)^2 + \rbsArrRate_{s}^2(t) \label{eq:lemma12}
\end{align}
where $(\mathrm{max}[\dataQueue_{su}(t),0])^2 \leq \dataQueue_{su}^2(t)$. Similarity squaring the virtual queue dynamics \eqref{eq:Deficit_queue1} and \eqref{eq:Deficit_queue2} and replacing $\dataQueue_{su}^f(t+1)^2$ and $\dataQueue_{cs}^f(t+1)^2$ respectively yields:
\vspace{-0.3cm}
\begin{align}
\textstyle \virtQueue_{su}^f(t+1)^2 & \ \leq \virtQueue_{su}^f(t)^2 + \Big(\epsilon_u \bar{\packetArrRate}_{su}^f \qosReq_{s\ueRequests_uf} - \dataQueue_{su}^f(t) + \sum_{m=1}^{\rbsSize}\rate_{su}^{(m,f)}(t) - \packetArrRate_{su}^f(t)\Big)^2 \nonumber
\end{align}
\begin{align}
& \textstyle \ + 2 \virtQueue_{su}^f(t) \Big(\dataQueue_{su}^f(t) - \sum_{m=1}^{\rbsSize}\rate_{su}^{(m,f)}(t) + \packetArrRate_{su}^f(t) - \epsilon_u \bar{\packetArrRate}_{su}^f \qosReq_{s\ueRequests_uf}\Big) \label{eq:lemma18} \\
\virtQueue_{cs}^f(t+1)^2 & \ \leq \virtQueue_{cs}^f(t)^2 + \Big(\frac {\epsilon_u {\qosReq_{sq_{uf}}}} {\packetSize_f} - \dataQueue_{cs}^f(t) + (1-\sbsTransDur) \frthaulPerUe_{s}^f \frthaulCap - \frac {1} {\packetSize_f} \sum_{u \in \ueCoverage_s(t)} \mathbbm{1}_{\{q_u(t) \not\in \insCache_s(t) \}}\Big)^2 \nonumber \\
& \textstyle \ + 2 \virtQueue_{cs}^f(t) \Big(\dataQueue_{cs}^f(t) - (1-\sbsTransDur) \frthaulPerUe_{s}^f \frthaulCap + \frac {1} {\packetSize_f} \sum_{u \in \ueCoverage_s(t)} \mathbbm{1}_{\{q_u(t) \not\in \insCache_s(t) \}} - \frac {\epsilon_u {\qosReq_{sq_{uf}}}} {\packetSize_f}\Big) \label{eq:lemma19} \end{align}
Summing \eqref{eq:lemma11}, \eqref{eq:lemma14} and \eqref{eq:lemma12} and  using definition of lyapunov drift, we have
\vspace{-0.3cm}
\begin{align*}
\textstyle \Delta(L) & (t) = L(t+1) - L(t) \\
\vspace{-0.4cm}
& \textstyle = \frac {1} {2} \Big( \packetArrRate_{su}^f(t) - \sum_{m=1}^{\rbsSize} \rate_{su}^{(m,f)}(t) \Big)^2 + \dataQueue_{su}^f(t) \Big( \packetArrRate_{su}^f(t) - \sum_{m=1}^{\rbsSize} \rate_{su}^{(m,f)}(t) \Big) + \frac {1} {2} \Big( \rbsArrRate_s(t) \nonumber \\
& \textstyle \ - \sum_{u \in \ueCoverage_s(t)} |\matchPolicy_{su}(t)| \Big)^2 + \rbQueueEvo_{s}(t) \Big( \rbsArrRate_s(t) - \sum_{u \in \ueCoverage_s(t)} |\matchPolicy_{su}(t)| \Big) + \frac {1} {2}
\Big( \frac {1} {\packetSize_f} \sum_{u \in \ueCoverage_s(t)} \mathbbm{1}_{\{q_u(t) \not\in \insCache_s(t) \} } \nonumber \\ 
& \textstyle \ - (1-\sbsTransDur) \frthaulPerUe_{s}^f \frthaulCap \Big)^2 + \dataQueue_{cs}^f(t) \Big( \frac {1} {\packetSize_f} \sum_{u \in \ueCoverage_s(t)} \mathbbm{1}_{\{q_u(t) \not\in \insCache_s(t) \} } - (1-\sbsTransDur) \frthaulPerUe_{s}^f \frthaulCap \Big) + \frac {1} {2} \Big( \epsilon_u \bar{\packetArrRate}_{su}^f \qosReq_{s\ueRequests_uf} \nonumber \\
& \ \textstyle - \epsilon_u \bar{\packetArrRate}_{su}^f \qosReq_{s\ueRequests_uf} \Big) + \frac {1} {2} \Big( \frac {\epsilon_s \qosReq_{s\ueRequests_uf}} {\packetSize_f} - \dataQueue_{cs}^f(t) + (1-\sbsTransDur) \frthaulPerUe_{s}^f \frthaulCap - \frac {1} {\packetSize_f} \sum_{u \in \ueCoverage_s(t)} \mathbbm{1}_{\{q_u(t) \not\in \insCache_s(t) \} } \Big)^2 \nonumber \\
& \ \textstyle + \virtQueue_{cs}^f(t) \Big( \dataQueue_{cs}^f(t) - (1-\sbsTransDur) \frthaulPerUe_{s}^f \frthaulCap + \frac {1} {\packetSize_f} \sum_{u \in \ueCoverage_s(t)} \mathbbm{1}_{\{q_u(t) \not\in \insCache_s(t) \}} - \frac {\epsilon_s \qosReq_{s\ueRequests_uf}} {\packetSize_f} \Big) \nonumber
\end{align*}
To ensure stability, the terms $\frac {1} {2} \Big( \packetArrRate_{su}^f(t) - \sum_{m=1}^{\rbsSet} \rate_{su}^{(m,f)}(t) \Big)^2$ for $\dataQueue_{su}^f(t)$, $\frac {1} {2} \Big( \rbsArrRate_s(t) - \\ \sum_{u \in \ueCoverage_s(t)} |\matchPolicy_{su}(t)| \Big)^2$ for $\rbQueueEvo_s(t)$, $\frac {1} {2} \Big( \frac {1} {\packetSize_f} \sum_{u \in \ueCoverage_s(t)} \mathbbm{1}_{\{q_u(t) \not\in \insCache_s(t) \} } - (1-\sbsTransDur) \frthaulPerUe_{s}^f \frthaulCap \Big)^2$ for $Q_{cs}^f(t)$, \\ $\frac {1} {2} \Big( \epsilon_u \bar{\packetArrRate}_{su}^f \qosReq_{s\ueRequests_uf} - \dataQueue_{su}^f(t) + \sum_{m=1}^{\rbsSet} \rate_{su}^{(m,f)}(t) - \packetArrRate_{su}^f(t) \Big)^2$ for $\virtQueue_{su}^f(t)$ and $\frac {1} {2} \Big( \frac {\epsilon_s \qosReq_{s\ueRequests_uf}} {\packetSize_f} - \dataQueue_{cs}^f(t) + \\ (1-\sbsTransDur) \frthaulPerUe_{s}^f \frthaulCap - \frac {1} {\packetSize_f} \sum_{u \in \ueCoverage_s(t)} \mathbbm{1}_{\{q_u(t) \not\in \insCache_s(t) \} } \Big)^2$ for $\virtQueue_{cs}^f(t)$ 
are required to be bounded. Therefore,
\vspace{-0.2cm}
\begin{align*}
\textstyle \Delta(L) \leq & \ C + \dataQueue_{su}^f(t) \Big( \packetArrRate_{su}^f(t) - \sum_{m=1}^{\rbsSize} \rate_{su}^{(m,f)}(t) \Big)  + \dataQueue_{cs}^f(t) \Big( \frac {1} {\packetSize_f} \sum_{u \in \ueCoverage_s(t)} \mathbbm{1}_{\{q_u(t) \not\in \insCache_s(t) \} } - (1-\sbsTransDur) \frthaulPerUe_{s}^f \frthaulCap \Big) \nonumber \\
& \textstyle + \rbQueueEvo_{su}(t) \Big( \rbsArrRate_s(t) - \sum_{u \in \ueCoverage_s(t)} |\matchPolicy_{su}(t)| \Big) + \virtQueue_{su}^f(t) \Big( \dataQueue_{su}^f(t) - \sum_{m=1}^{\rbsSize} \rate_{su}^{(m,f)}(t) + \packetArrRate_{su}^f(t)  \nonumber \\
& \textstyle - \epsilon_u \bar{\packetArrRate}_{su}^f \qosReq_{s\ueRequests_uf} \Big) + \virtQueue_{cs}^f(t) \Big( \dataQueue_{cs}^f(t) - (1-\sbsTransDur) \frthaulPerUe_{s}^f \frthaulCap + \frac {1} {\packetSize_f} \sum_{u \in \ueCoverage_s(t)} \mathbbm{1}_{\{q_u(t) \not\in \insCache_s(t) \}} - \frac {\epsilon_s \qosReq_{s\ueRequests_uf}} {\packetSize_f} \Big)
\end{align*}
where $C$ is the uniform bound on $\frac {1} {2} \Big[ \Big( \packetArrRate_{su}^f(t) - \sum_{m=1}^{\rbsSize} \rate_{su}^{(m,f)}(t) \Big)^2 + \Big( \rbsArrRate_s(t) - \sum_{u \in \ueCoverage_s(t)} |\matchPolicy_{su}(t)| \Big)^2 + \Big( \frac {1} {\packetSize_f} \sum_{u \in \ueCoverage_s(t)} \mathbbm{1}_{\{q_u(t) \not\in \insCache_s(t) \} } - (1-\sbsTransDur) \frthaulPerUe_{s}^f \frthaulCap \Big)^2 + \Big( \epsilon_u \bar{\packetArrRate}_{su}^f \qosReq_{s\ueRequests_uf} - \dataQueue_{su}^f(t) + \sum_{m=1}^{\rbsSize} \rate_{su}^{(m,f)}(t) - \packetArrRate_{su}^f(t) \Big)^2 + \Big( \frac {\epsilon_s \qosReq_{s\ueRequests_uf}} {\packetSize_f} - \dataQueue_{cs}^f(t) + (1-\sbsTransDur) \frthaulPerUe_{s}^f \frthaulCap - \frac {1} {\packetSize_f} \sum_{u \in \ueCoverage_s(t)} \mathbbm{1}_{\{q_u(t) \not\in \insCache_s(t) \} } \Big)^2 \Big]$